\documentclass[preprint,12pt]{elsarticle}

\usepackage{amssymb}
\usepackage{booktabs}
\usepackage{multirow}
\usepackage{url}

\journal{Computer science}

\begin{document}

\begin{frontmatter}



\title{A Distributed Blockchain-based Access Control for
the Internet of Things}


\author[inst1]{Ebtihal Abdulrahman}

\affiliation[inst1]{organization={Department of Information Technology, King Abdulaziz University},
            addressline={iabdulrahman0001@stu.kau.edu.sa}, 
            city={Jeddah},
            country={Saudi Arabia }}

\author[inst2]{Suhair Alshehri}

\affiliation[inst2]{organization={Department of Information Technology, King Abdulaziz University},
            addressline={sdalshehri@kau.edu.sa}, 
            city={Jeddah},
            country={Saudi Arabia}}

 \author[inst4]{Ali Alzubaidy}            

 \affiliation[inst4]{organization={Department of Computer Science, Umm Al-Qura University},
            addressline={aakzubaidi@uqu.edu.sa}, 
            city={Makkah},
            country={Saudi Arabia}}

 \author[inst3]{Asma Cherif}

 \affiliation[inst3]{organization={Department of Information Technology and Center of Excellence in Smart Environment Research, King Abdulaziz University},
            addressline={acherif@kau.edu.sa}, 
            city={Jeddah},
            country={Saudi Arabia}}

\begin{abstract}
Recently, the Internet of Things (IoT) environment has become increasingly fertile
for malicious users to break the security and privacy of IoT users. Access control is a
paramount necessity to forestall illicit access. Traditional access control mechanisms
are designed and managed in a centralized manner, thus rendering them unfit for
decentralized IoT systems. To address the distributed IoT environment, blockchain
is viewed as a promising decentralised data management technology. In this thesis,
we investigate the state-of-art works in the domain of distributed blockchain-based
access control. We establish the most important requirements and assess related works against them. We propose a Distributed Blockchain and Attribute-based Access Control model for IoT entitled (DBC-ABAC) that merges blockchain technology
with the attribute-based access control model. A proof-of-concept implementation
is presented using Hyperledger Fabric. To validate performance, we experimentally
evaluate and compare our work with other recent works using Hyperledger Caliper
tool. Results indicate that the proposed model surpasses other works in terms of
latency and throughput with considerable efficiency.
\end{abstract}




\end{frontmatter}


\section{Introduction}
In recent years, researchers and developers have applied prodigious efforts to the Internet of Things (IoT) domain, resulting in an exponential increase in the number of IoT devices connected to the Internet. It is projected that these devices will reach 75 billion by 2025 and 500 billion by 2030 \cite{ref2}. However, IoT devices are prone to collecting sensitive data from the surrounding environment and sharing it with other entities. This data is susceptible to violation by malicious users, thereby raising significant security concerns.   

For example, a famous attack was carried out to steal sensitive data from NASA with the use of a portable and economical Raspberry Pi device \cite{ref36}.  Therefore, it is imperative to equip IoT systems with robust yet efficient access control models to prevent unauthorized access, thus preserving sensitive data. In large-scale IoT systems such as smart cities or with the mobility feature of some IoT devices as in Vehicle Ad-hoc Network (VANET), traditional access control is rendered inefficient owing to its centralized nature \cite{ref3}.

Since the IoT architecture recently moved from centralization to decentralization in terms of management operations (i.e., device identification, authentication, and authorization) \cite{ref5}, researchers have started to work on decentralized access control approaches 
since they provide better scalability, more efficient security, privacy, and reliability. 

\begin{figure}[htp!]
    \centering
    \includegraphics[scale=.33]{  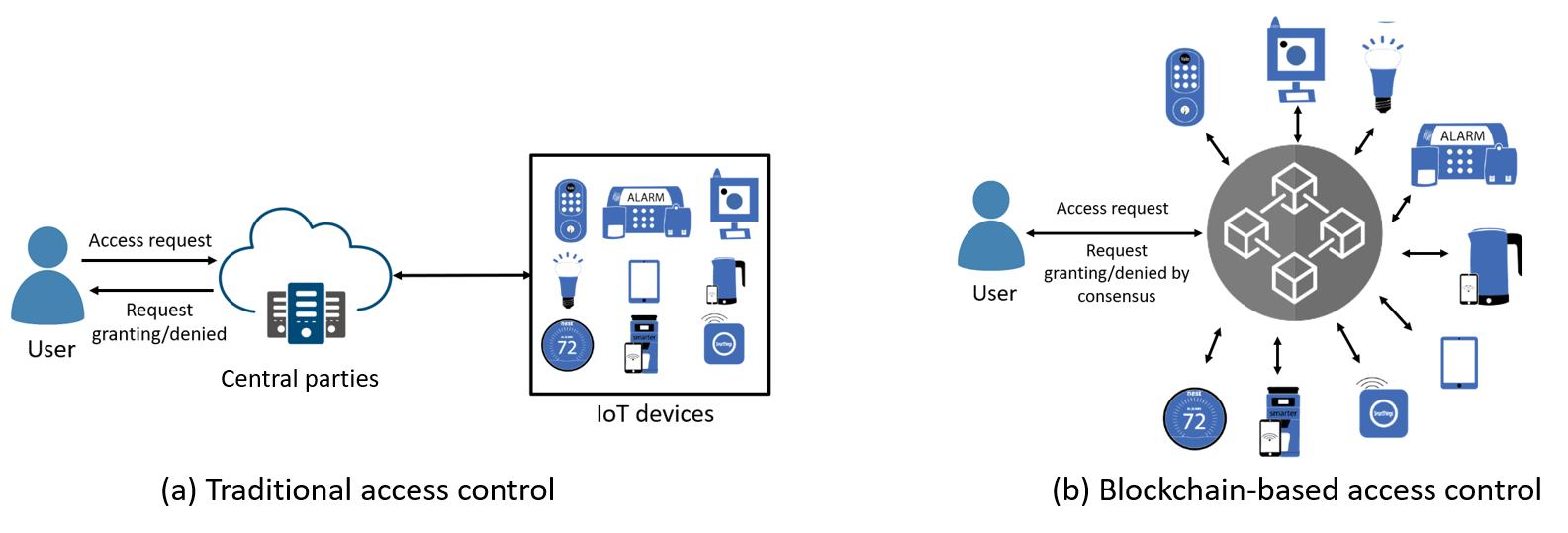}
    \caption{(a,b) Traditional vs blockchain-based access control}
    \label{fig:motiv}
\end{figure}

Blockchain is one of the promising decentralized systems that attracted researchers in the domain of IoT access management due to distributed ledgers (see Figure~\ref{fig:motiv}), immutability (i.e., once an entry is validated, it cannot be modified or deleted.), traceability, and information sharing and management without needing any third party \cite{ref3}. 

\section{Problem Definition and Motivation}

Applying access control to the IoT environment is a challenging problem due to the characteristics of IoT devices.

Various  centralized access control models such as Mandatory Access Control (MAC), Discretionary Access Control (DAC),  Role-Based Access Control (RBAC), and Attribute-Based Access Control (ABAC) exist, yet these models suffer from certain security and privacy limitations, rendering them inadequate for IoT access management; for example:

\begin{itemize}

\item Centralized Management: These models rely upon a central administrator to manage the operations between entities (e.g., user identification, role and group definitions) \cite{ref2}. However, the centralized nature of traditional access control models leads to a single point of failure issue \cite{ref3,ref4,ref18}.

\item Centralized Storage: Most traditional access models store all policies in a centralized repository, raising concerns regarding privacy and security \cite{ref2,ref3,ref5,ref20}.
\item Scalability: Limited scalability impedes the ability to accommodate the exponentially growing number of IoT devices \cite{ref3,ref4}.

\item Resource consumption: Intensive computation related to access management renders it incompatible with resource-limited IoT devices \cite{ref3,ref4}.

\item Static:  Static nature of certain access control models disqualifies them from accommodating the mobility characteristics of some IoT devices \cite{ref3,ref4,ref20}.

\end{itemize}




Blockchain technology can be utilized to enhance access control by providing an immutable, secure, and distributed ledger that can store access control policies, facilitate scalability, and support mobility \cite{ref1,ref2}. Nevertheless, certain limitations have been identified in the existing applications of blockchain technology with regard to the security of IoT networks, such as:

\begin{itemize}

\item Privacy leakage: Certain blockchain characteristics, such as auditability and transparency, facilitate the tracking of transactions and analysis of user behavior by a third party, potentially leading to significant privacy issues in many IoT systems. \cite{ref3, ref34}.


\item Throughput: Thousands of transactions need to be handled per second in a scaled IoT network \cite{ref1}. However, traditional blockchain models are unable to process such high volumes of transactions. For example, Bitcoin and Ethereum have limited throughputs, with Bitcoin capable of handling only 3-7 transactions per second and Ethereum only 20 transactions per second, making them inadequate for the needs of the IoT \cite{ref9}.

\item Latency:  Verification of transactions and generation of blocks in the blockchain system results in a non-trivial delay, even with a limited transaction rate in relatively small IoT infrastructures (i.e., small office environments) \cite{ref1}. For instance, Bitcoin requires 30 minutes to confirm each transaction, making it inefficient for IoT systems that require more stringent latency constraints \cite{ref9}.

\item Overhead and scalability: In the standard blockchain system, all nodes verify all broadcasted blocks, resulting in major scalability and overhead issues. The processing overhead and broadcast traffic increase linearly with the increasing number of connected nodes. This is further exacerbated by the limited processing capabilities and bandwidth connections of most IoT devices \cite{ref9}.

\item Complex consensus algorithms: High resource consumption is a significant challenge with several of the existing consensus algorithms, making them incompatible with most IoT devices \cite{ref2,ref9}. Furthermore, certain consensus algorithms necessitate an abundance of electricity to compute the hash, and this demand increases proportionally with the expansion of an IoT network \cite{ref17}.
\end{itemize}

Recently, some research efforts have been directed toward applying blockchain to enforce and manage access control in a distributed fashion. Unfortunately, the current state-of-the-art is still in its early stage. 
Though some blockchain-based access control models have been proposed, they still rely on heavy consensus algorithms or incur additional overhead due to heavy communication between nodes \cite{ref5,ref16}. Furthermore, some of these works still rely on a central trusted party (i.e, cloud) \cite{ref7} or require additional burden on the resource-constrained IoT device to make the access decision inside them \cite{ref16}. Also, most of the proposed models do not take into account the access delegation operation to transfer access rights between network entities and offer better flexibility and more decentralized access management. These  gaps raise the following research questions:


\begin{itemize}

\item \textbf{RQ1}: How can a distributed access control model be designed utilizing blockchain technology?
\item \textbf{RQ2}: How can blockchain limitations be mitigated while taking IoT constraints into account?

\end{itemize}

To answer the above questions, the purpose of this thesis is to (1) identify a set of requirements for a distributed blockchain-based access control model for IoT systems and (2) design and implement a distributed blockchain-based access control model considering the defined requirements.

\section{Research Objectives}

The main research objectives are stated as follows:

\begin{itemize}

\item \textbf{OBJ1}: To design a distributed blockchain-based access control model for IoT.

\item \textbf{OBJ2}: To support the delegation of the access rights. 
\item \textbf{OBJ3}: To achieve high efficiency in terms of latency and throughput.

\end{itemize}

\subsection{Contributions}

 In this thesis, we investigate the most significant requirements to devise and implement an efficient distributed blockchain-based access control model for IoT environments. The contributions of our thesis are as follows:

\begin{enumerate}
\item  Identify the most significant requirements for constructing an efficient distributed blockchain-based access control model for IoT environments.
\item   Design a distributed blockchain-based access control model that satisfies the identified requirements.

\item   Improve performance in terms of latency and throughput to manage extensive transactions in a scalable IoT environment.

\item   Support access right delegation between network entities.
\end{enumerate}

\subsection{Paper Organization}
This research paper is structured as follows: Section \ref{fig:LR} reviews current literature and compares the state of the art against a well-established set of requirements. Section \ref{sec:Sol} presents the theoretical framework, enhanced security protocols, and risk assessment methodology. Implementation and Testing details the experimental setup, testing procedures, and data collection methods. Section \ref{sec:test} presents the testing results and their discussion. Finally, Section \ref{sec:conc} summarizes key findings, outlines contributions to the field, and suggests future research directions.  

\section{Related Works} \label{sec:RW}

\begin{figure}[ht]
    \centering
    \includegraphics[scale=0.49]{   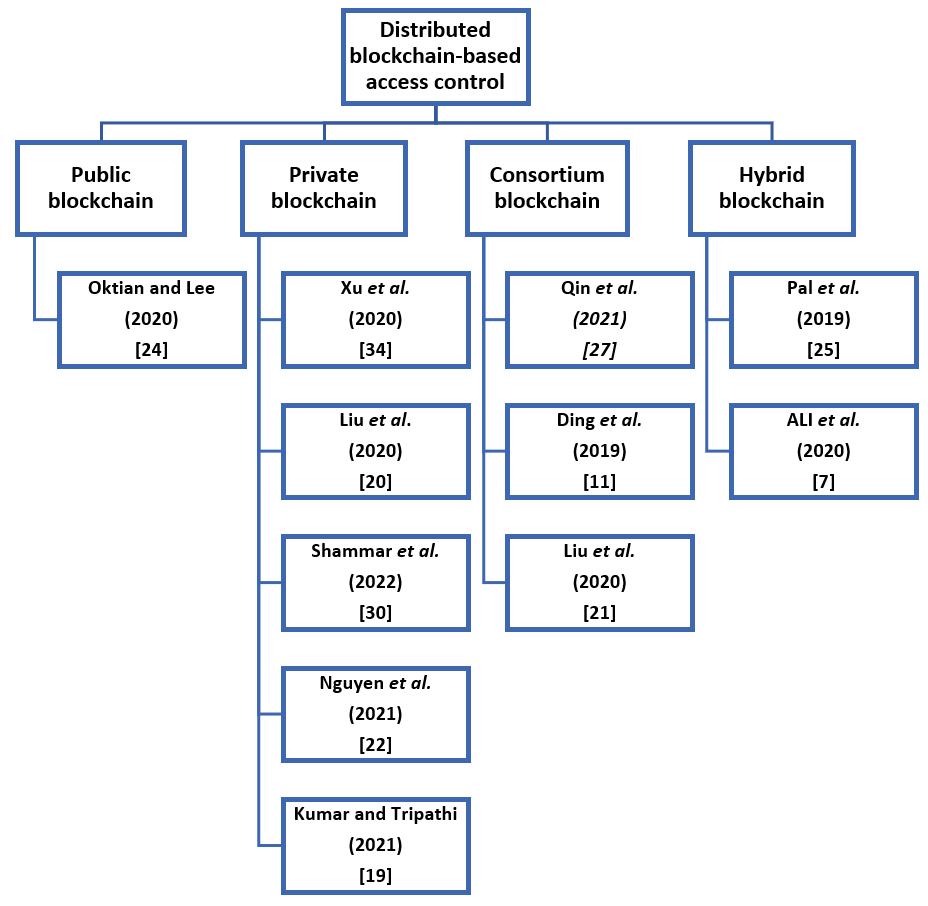}
    \caption{Taxonomy of the related works}
    \label{fig:LR}
\end{figure}

Several works have been suggested in the domain of distributed blockchain-based access control for the Internet of Things. In this chapter, we investigate the most recent models based on the taxonomy illustrated in Figure ~\ref{fig:LR}. In the following, we review the main proposed models for each blockchain type. Finally, we define a set of requirements and compare the reviewed works against it.

\subsection{Public blockchain-based models}

Oktian and Lee \cite{ref5} used the Ethereum blockchain system to provide an access control framework for IoT endpoints named BorderChain. It allows the IoT domain owner to control the authentication and authorization processes for his endpoint. After authenticating the location for each gateway (GW) and the identity for all devices through central trusted approvers, only authorized access requests are permitted by the domain owner. All authentication and authorization tokens are activated and stored in the trusted list within the smart contract. Then, activated access tokens are used by the IoT services for access requests or revocation procedures. However, centralization is the main issue in Borderchain due to the use of trusted approvers. Also, implementing the Ethereum blockchain with PoW using NodeJS leads to limited scalability and inefficiency in terms of latency and throughput. Security and privacy problems also arise due to the use of a public blockchain.

\subsection{Private blockchain-based models}

Xu et al. \cite{ref12} proposed a blockchain-based secure data-sharing platform with fine-grained access control (BSDS-FA) to solve privacy leakage problems during data sharing. This model supports one-to-many encryption scenarios by suggesting a novel hierarchical attribute-based encryption algorithm (HABE). The Fabric blockchain technology combined with HABE to enhance the performance by deploying two different smart contracts which are the Validation Contract and Decryption Contract. For proper management and to enhance the performance of each authorization center, various user attributes are distributed to various authorization agencies which provide fine-grained and flexible access control. The model is fully decentralized but it raised computing overhead on the user side. The user needs to apply part of the encryption to obtain shared data.

Liu et al. \cite{ref7} designed and implemented an open-source Fabric-IoT access control system that combines attributed-based access control (ABAC) with the Hyperledger Fabric blockchain framework. This solution provides dynamic management of permissions, a decentralized architecture to separate devices and users, and fine-grained access control. To simplify device resources’ storage structure and sharing model, the Fabric-IoT allows the device to grant access to the data resources based on a URL. Three different smart contract applications are implemented to ensure access security of the device resources which are: device, policy, and access contract. However, this model needs to improve its scalability to support more IoT application integration. 

Shammar et al. \cite{ref28} enhanced the Fabric-IoT \cite{ref7} by producing a distributed and lightweight attribute-based access control model using the Hyperledger Fabric blockchain (ABAC-HLFBC) model. ABAC-HLFBC used the Hyperledger Fabric Raft consensus mechanism for transaction verification rather than Kafka ordering service. Unlike Kafka, Raft is less complicated and faster. The evaluation results proved that the proposed model outperforms the Fabric-IoT regarding latency and throughput. Utilizing cloud service as external data storage in this work raises a centralization problem.

Kumar and Tripathi \cite{ref20} addressed the scalability problem in blockchain technology and improved the dynamic behavior of the access control model by employing an enhanced Bell-Lapadula (EBLP) model integrated with fabric blockchain for the healthcare system. They used smart contracts to design novel access control policies (read, remark, update, and delete), thus supporting dynamic behavior. Each node keeps the state of the ledger according to its access control policy, thus enhancing scalability. Also, the clearance and security level of transactions and peers help to improve security and privacy issues. EBLP does not rely on any third parties which makes it a  decentralized model.

A novel decentralized BEdgeHealth model proposed by Nguyen et al.\cite{ref32} integrated the blockchain and mobile edge computing (MEC) for controlling access to health data records in a distributed health industry. Two schemes were developed for data offloading and data sharing. In a privacy-aware data offloading scheme, the IoT health data was offloaded to the nearest MEC to reduce the computational overhead on the IoT devices.  A cooperation of blockchain, MEC, distributed interplanetary file system (IPFS), and smart contracts are used to implement access control in the data sharing scheme. Access contract smart contract deployed at network edge without needing any third party which reduces access control latency. The effectiveness of the proposed model in terms of security and performance is demonstrated by a real-world experiment.

\subsection{Consortium blockchain-based models}

Qin et al. \cite{ref10} provided a lightweight blockchain-based access control scheme (LBAC) that is based on both attribute-based encryption (ABE) and Hyperledger Fabric \cite{ref13} blockchain technology. The proposed scheme solved the expensive operations problem caused by ABE by outsourcing the decryption operation to the blockchain. They designed four smart contracts to eliminate any extra validation overhead on the user side.  Also, to guarantee the correctness of the outsourced decryption. LBAC assumed that the cloud is untrusted to avoid user deception. Furthermore, the endorsement protocol is dynamically adjusted after measuring the user’s credibility based on the user’s abnormal access. However, this model is not fully decentralized due to the use of  the cloud and attribute authorities in the architecture. Also, some transactions cause high latency and low throughput. Dynamic management of the attributes needs to be considered in the future.

To simplify access management, Ding et al. \cite{ref18} produced a new attribute-based access control scheme that uses a set of predefined attributes generated by the attribute authorities to describe each device based on the identity. They utilized the blockchain to record the distribution of the attributes to mitigate the centralization issue. These attributes are used by the data owner to authorize the data requester before exchanging data. The model also supports attribute revocation to provide dynamic management.  Security analysis proves the resistance against various attacks. However, the model incurs some computation overhead due to the extra verification calculations performed by the data consumer.

Liu et al. \cite{ref22} used the blockchain jointly with Decentralized Identifiers (DID). They suggested a decentralized capability-based IoT model where at least one DID identifies each participant in the IoT network. The access control for a specific device is restricted by credentials and the capability/ownership token. The architecture model is composed of three different capability-based IoT access control modules, and each module is associated with an on-chain smart contract that performs various operations. However, the model relies on edge servers to manage the IoT devices, making access management centralized (i.e., vulnerable to a single point of failure). Furthermore, the response time of some critical services is high, which is not suitable in some IoT scenarios.

\subsection{Hybrid blockchain-based models}

Pal et al. \cite{ref16} proposed a flexible and scalable access right delegation model for IoT using both public and private Ethereum blockchain. The proposed model mitigates the privacy issue by moving the critical attributes from the public blockchain into the private one while the public blockchain carries other computations. Two different designs were proposed, the first one is more secure, but not suitable for distributed IoT environments due to centralized access to the attributes. The second is   decentralized  but   raises  security and performance issues because the attributes can be requested by any contract with delegation permissions and the long access confirmation time which is a limiting factor when moving the access to the IoT device.

To support the virtual coalition, the xDBAuth framework is proposed by ALI et al. \cite{ref23} that is based on a hierarchy of local and global smart contracts. The framework performs access control and delegation within a particular domain and across different domains for internal and external users/IoT devices. The delegation policies are stored in the MongoDB off-chain storage to overcome the limited storage on the blockchain. However, the total number of delegation policies was restricted due to the limited scalability of MongoDB.

Table~\ref{tab:summary} summarises the proposed model and its main characteristics such as the access control model, blockchain platform, blockchain type, and consensus algorithm of each proposed work surveyed in the literature review.

\begin{table}[htbp]
\begin{scriptsize}
    \centering
    \caption{Summary of related works}
    \label{tab:summary}
    \begin{tabular}{p{0.5cm} p{0.7cm} p{3cm} p{3cm}   p{2cm} p{1.5cm} p{1.5cm}}

    \textbf{Ref.} &	\textbf{Year}& \textbf{Proposed Model}	& \textbf{Access Control}
\textbf{Model}&	\textbf{Blockchain
Platform}&	\textbf{Blockchain 
Type}&	\textbf{Consensus
Algorithm}
\\ \toprule

\cite{ref10}&
2021&
Lightweight blockchain-based
access control scheme (LBAC)&	Ciphertext-policy attribute-based encryption (CP-ABE)	&
Hyperledger Fabric	&
Consortium	&
Solo\\\\

\cite{ref20}&2021&	Enhanced Bell-LaPadula (EBLP)
model and blockchain integration	Enhanced &Bell–LaPadula model
(EBLP)	&Hyperledger Composer&	Private	&NA
\\\\

\cite{ref5}	&2020 &BorderChain & Identity-based access control &	Ethereum	&Public&	PoW
\\\\

\cite{ref7}	&2020& Fabric-IoT&	Attribute-based access control (ABAC)	& Hyperledger Fabric&	Private	&Kafka
\\\\

\cite{ref16}&2019&	Access right delegation model using blockchain technology	& Attribute-based access control (ABAC)	& Ethereum&	Hybrid	& PoW
\\\\

\cite{ref12}	&2020 &Blockchain-Based Data Security
Sharing Platform with Fine-Grained
Access Control
(BDSS-FA)	&Hierarchical attribute-based 
encryption algorithm
(HABE)	&Hyperledger Fabric&	Private	&Kafka
\\\\

\cite{ref18}&2019 &	A Novel Attribute-Based Access Control
Scheme Using Blockchain for IoT&	Novel attribute-based access control & 	Hyperledger Fabric	&Consortium	&PBFT
\\\\

\cite{ref22}&
2020&
Capability-based IoT access control using blockchain&
Capability-based access control (CBAC)&
Ethereum	&
Consortium	&
PoA
\\\\

\cite{ref23}&
2020&
xDBAuth&
Attribute-based encryption (ABE)&
They build their own chain using Node.js	&
Hybrid	&
PoAI
\\\\

\cite{ref32}	&2021 &BEdgeHealth
	& Identity-based access control 
	&Hyperledger Fabric&	Private	&PBFT

\\\\

\cite{ref28}	&2022 &ABAC-HLFBC
	& Attribute-based access control (ABAC)
	&Hyperledger Fabric&	Private	&Raft

\\
\bottomrule
\end{tabular}
\end{scriptsize}
\end{table}

\subsection{Comparing distributed blockchain-based access control models}

After presenting several models that provide distributed blockchain-based access control for IoT. We defined the most important requirements \cite{ebtihel} that must be considered while designing a distributed blockchain-based access control model for IoT. Then, we compare these models against the defined requirements. In the following, we list and define these requirements.

\begin{itemize}
\item Lightweight: The model should preserve low computational, communicational, and storage overhead due to the constrained resources of IoT devices \cite{ref3}.

\item Scalability: The model should handle the growing number of IoT devices \cite{ref3}.

\item Dynamicity: The model should provide dynamicity in terms of flexible  policy management \cite{ref3}.

\item Decentralized: The model should be decentralized in terms of architecture and data storage.

\item Latency: The model should maintain real-time response for most IoT systems \cite{ref4}.

\item Throughput: The model should handle a high number of transactions in the scaled IoT network \cite{ref1}.

\item Delegation: The model should support the transfer of access rights between entities in the IoT network \cite{ref16}.

\item Revocation: The model should repeal the authorization decisions when needed \cite{ref5,ref18}.
\end{itemize}

In Table~\ref{tab:comparaison}, we compare the existing reviewed access control mechanisms against the identified requirements to identify the gap.

\begin{table*}[h]
\begin{scriptsize}

\centering 
    \caption{Comparing existing works}
    \label{tab:comparaison}
    \begin{tabular}{p{1.5cm}p{1.7cm} p{0.4cm} p{0.4cm} p{0.4cm} p{0.4cm} p{0.4cm} p{0.4cm} p{0.4cm} p{0.4cm} p{0.4cm} p{0.4cm} p{0.4cm}}
\multicolumn{2}{c}{Requirements} &	 

\multicolumn{1}{l}{\cite{ref10}}&	 
\multicolumn{1}{l}{\cite{ref20}} &
\multicolumn{1}{l}{\cite{ref5}}	&
\multicolumn{1}{l}{\cite{ref7}}	& 
\multicolumn{1}{l}{\cite{ref16}}&
\multicolumn{1}{l}{\cite{ref12}}&
\multicolumn{1}{l}{\cite{ref18}} &
\multicolumn{1}{l}{\cite{ref22}} &
\multicolumn{1}{l}{\cite{ref23}} &
\multicolumn{1}{l}{\cite{ref32}} &
\multicolumn{1}{l}{\cite{ref28}} 
\\ \toprule

\multirow{3}{*}{Lightweight}& Computation	 & $\times$	&\checkmark  & $\times$	 &	\checkmark  &	\checkmark  &$\times$	 &\checkmark &\checkmark & $\times$	  &\checkmark  &\checkmark \\ 

& Communication	&$\times$ 	&$\times$ &\checkmark	 &\checkmark	 &$\times$	 &\checkmark 	 &\checkmark &$\times$ &$\times$	 &\checkmark  &\checkmark \\ 

& Storage	&\checkmark &$\times$ &\checkmark	 &\checkmark	 &$\times$	 
&\checkmark 	 &\checkmark	&\checkmark &\checkmark  &\checkmark  &\checkmark\\  \hline

\multirow{2}{*}{Decentralized} &Architecture &$\times$ &\checkmark &$\times$  &\checkmark &$\times$ &$\times$ &\checkmark &$\times$ &\checkmark &\checkmark &\checkmark \\ 

&Storage &$\times$ &\checkmark &$\times$ &$\times$ &\checkmark &$\times$ &$\times$  &\checkmark &$\times$  &\checkmark &$\times$\\  \hline

\multicolumn{2}{l}{Scalability}	&\checkmark &\checkmark &$\times$	&\checkmark &$\times$ &\checkmark	 &\checkmark	 &$\times$ 	&\checkmark &\checkmark 		&\checkmark  \\

\multicolumn{2}{l}{Dynamicity}	&\checkmark	&\checkmark&	\checkmark&	\checkmark&	\checkmark&	\checkmark&	\checkmark &	\checkmark &	\checkmark &	\checkmark &	\checkmark \\

\multicolumn{2}{l}{Latency}	&$\times$ &\checkmark &$\times$ &\checkmark	&$\times$	&\checkmark	&$\times$ &$\times$ 	&\checkmark &\checkmark &\checkmark\\

\multicolumn{2}{l}{Throughput}&$\times$	&$\times$ &$\times$ &\checkmark &$\times$ &\checkmark &$\times$ &$\times$  &\checkmark &\checkmark &\checkmark\\

\multicolumn{2}{l}{Delegation} &$\times$	  &$\times$	  &$\times$	  &$\times$	  &\checkmark &\checkmark &$\times$ &$\times$	  &\checkmark &$\times$ &$\times$\\

\multicolumn{2}{l}{Revocation}&	$\times$ 	&$\times$ 	&\checkmark&	\checkmark &$\times$	 &$\times$	 &	\checkmark &	\checkmark&	\checkmark&$\times$  &\checkmark	 \\

\bottomrule
\end{tabular} 
    
\end{scriptsize}
\end{table*}

 The comparison table above reveals that none of the reviewed models is satisfying all of the given requirements. We point out that works \cite{ref7}, \cite{ref32}, and \cite{ref28} satisfy most of the predefined requirements. Accordingly, these exemplary works served as an inspiration to address the research gap. Although Fabric-IoT \cite{ref7} and ABAC-HLFBC \cite{ref28} achieve most requirements, they still have some limitations in terms of centralized storage due to the use of the cloud to store IoT data. Also, they do not support delegation in their models. Nevertheless, they do boast an advantage related to their access control model, which is the use of attribute-based access control (ABAC). The third most noteworthy work, BEdgeHealth \cite{ref32}, surpasses its predecessors in terms of distributed architecture and storage; however, it falls short in that it does not support either revocation or delegation. To come up with our model, we inspire by the advantages of the best works to fill the research gap in the comparative study which are the distributed architecture and storage from BEdgeHealth \cite{ref32} and the attribute-based access control (ABAC) model from Fabric-IoT \cite{ref7} and ABAC-HLFBC \cite{ref28}.

\section{Proposed Model}\label{sec:Sol}

Drawing inspiration from the three preeminent works identified in the previous section \cite{ref7}, \cite{ref32}, and \cite{ref28}, this section presents our model. Section 3.1 introduces a Distributed Blockchain and Attribute-based Access Control (DBC-ABAC) model for the Internet of Things.  Section 3.2 outlines the overall access control model and data access policy model.  Section 3.3 delineates how we configure smart contracts to disseminate access control into the blockchain.  Section 3.4 shows how we combine the access control with the proposed architecture. Finally, we explain the system workflow in section 3.5.

\subsection{DBC-ABAC Model}

This section illustrates the proposed system architecture and domain-layered view. Additionally, it elucidates the attribute-based access control (ABAC) model and how its components are situated within our architecture. Furthermore, we provide the ABAC data access policy model and finally the design of the smart contracts.

\subsubsection {System Architecture}

\begin{figure}[htp!]
    \centering
    \includegraphics[scale=.32]{ 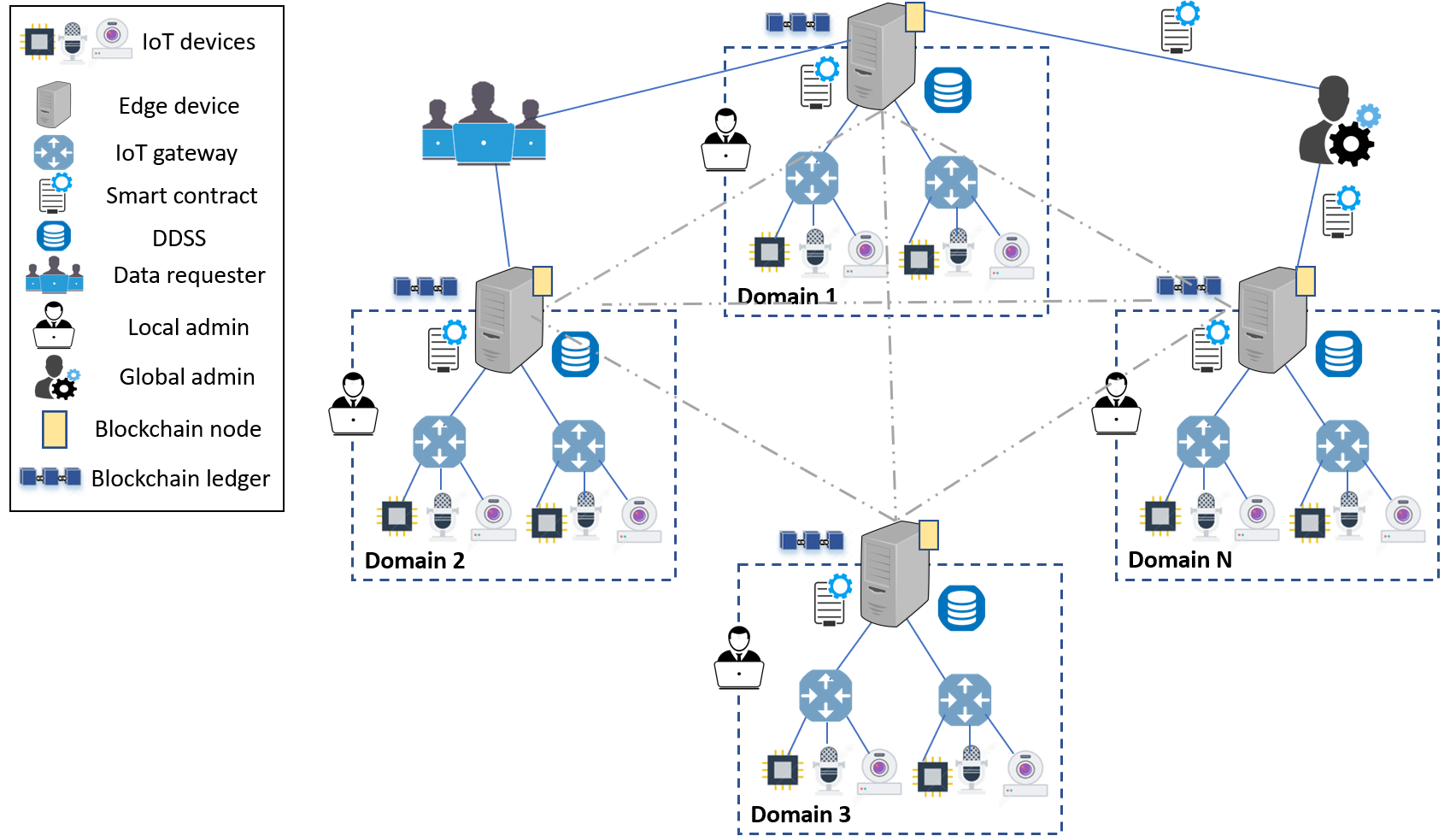}
    \caption{system architecture}
    \label{fig1:architecture}
\end{figure}

Based on \cite{ref32}, the proposed architecture depicted in Figure~\ref{fig1:architecture} consists of a network of cooperative domains interconnected by a blockchain ledger. Each domain is managed by an edge device that utilizes peer-to-peer (P2P) communication to interact and share data with other edge devices. A set of IoT gateways are located at each domain to collect data from a set of IoT devices and forward it to the edge. Additionally, we consider a set of users who may be situated in any domain and seek to access IoT data on the edge network. The components of each domain are outlined as follows:

\begin{itemize}

    \item IoT devices: They are responsible for gathering  data. Given their resource-constrained nature, low energy and storage capacity, limited processing power, and short-range communication capability, it is unfeasible to deploy these IoT devices as a blockchain node directly \cite{ref7}. Therefore, the IoT device transmits the gathered data to the closest powerful device, such as an IoT gateway, for transmission to the blockchain.
    
    \item IoT gateways: They are responsible for acquiring raw data from IoT devices and transferring it to the blockchain. Serving as a gateway between the IoT devices and the blockchain, they help avert the blockchain system from being inundated by direct access from IoT devices \cite{ref7}. Furthermore, IoT gateways have greater processing power and battery life in comparison to IoT devices, making them more suitable for deployment as blockchain nodes.
    
    \item Edge devices: These are blockchain nodes   responsible for managing a cluster of IoT gateways in its domain to provide low-latency computation services. Each edge node links with the other edge nodes from other domains in a peer-to-peer (P2P) fashion to construct a decentralized data-sharing network. Taking into account real-world scenarios, the edge node may be semi-trusted, implying that it may be inquisitive about collected data. To confront this challenge, we store data in off-chain storage (DDSS) while only the hash value of data is retained in the smart contract so that any data retrieval actions are traced on the blockchain in a transparent manner.
    
    \item Distributed Data Storage System (DDSS):   As the edge nodes are semi-trusted, data collected from IoT devices can be uploaded to a Distributed Data Storage System (DDSS), such as InterPlanetary File System (IPFS), in order to overcome the storage limitations of the blockchain and enhance overall performance with efficient P2P data delivery, fast decentralized archiving, and high-throughput block storage \cite{ref33}, \cite{ref32}.

    \item Smart contract: Each edge node holds a copy of the smart contract which is the core component of our model. It is responsible for managing access operations and access policies.

    \item User: the user is classified into two types in this architecture: admin, and normal user.

    \begin{enumerate}
    
        \item Admin: We distinguish between two types of blockchain admins:  global and local.

    \subsubitem - Global admin: The global administrator is responsible for managing the entire blockchain network. The specific operations allowed are adding and upgrading smart contracts.
    
    \subsubitem - Local admin: The administrator of each domain is responsible for managing the access policies of users and IoT data located therein. These policies can be added, queried, updated, and deleted. The local administrator also has the power to revoke access authorization in the event of any violations. 
    
        \item Normal user: The "normal user" refers to the data requester who wishes to gain access to the data record; this could be either the data owner or another user. The data owner is able to delegate access rights over their owned data record to other users.  This work utilizes the terms "normal user" and "data requester" interchangeably.
        
    \end{enumerate}

\end{itemize}

\subsubsection{Domain Layered View}

\begin{figure}[htp!]
    \centering
    \includegraphics[scale=0.46] {    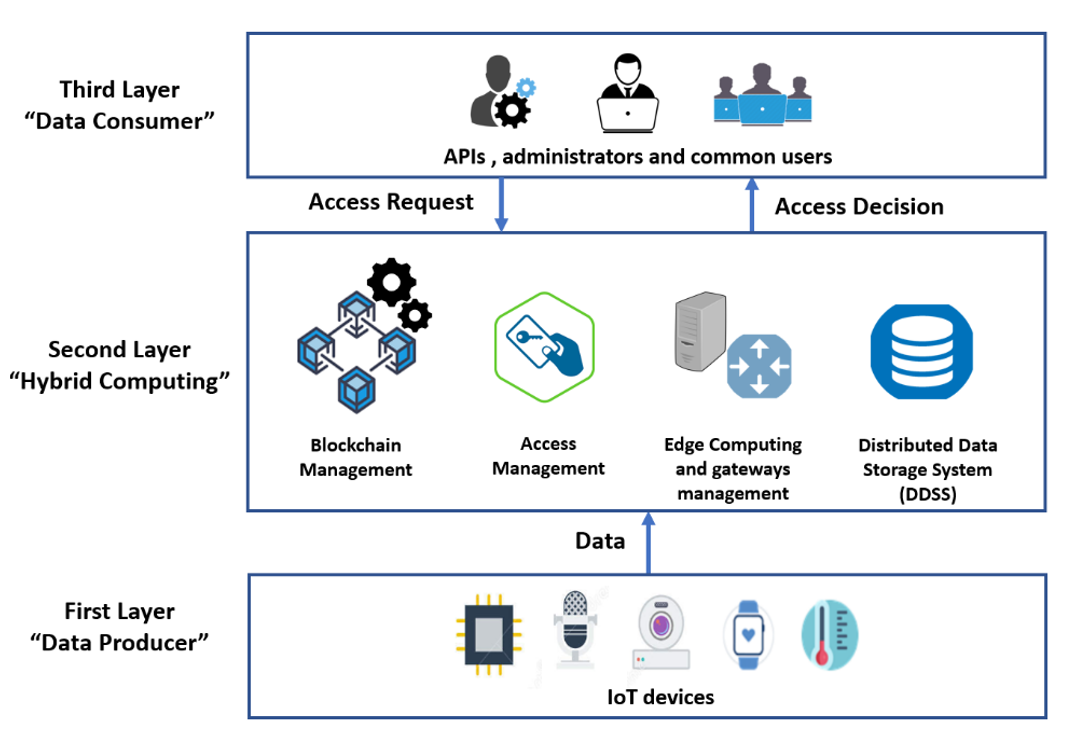}
    \caption{Domain layered view}
    \label{fig1:layered}
\end{figure}

 It is assumed that each domain has a three-layered architecture comprising of a data producer layer, hybrid computing layer, and data consumer layer (see Figure~\ref{fig1:layered}).

\begin{itemize}

    \item Data producer layer: This layer consists of IoT devices that collect data from their surrounding environment and transmit it to the second layer. 
    
    \item Hybrid computing Layer: This layer combines both edge computing and blockchain to gain the benefits of both technologies, while also containing the off-chain Distributed Data Storage System (DDSS) for storing collected data.  acts as a bridge between the IoT devices and the blockchain to relieve pressure on the blockchain system, while access control is managed by this layer as well.
    
    \item Data consumer layer: This layer houses the Application Programming Interfaces (APIs) used by administrators and users to request access, as well as receive the access decision returned by the middle layer.
    
\end{itemize}

\subsection{Access Control Specification}

Given the requirements previously outlined in Chapter 3 to implement an effective blockchain-based access control model for IoT systems, we opted to apply the Attribute-Based Access Control (ABAC) model and combine it with the proposed architectural model. Indeed, compared to other access control models, ABAC is the most suitable choice for controlling access over sensitive data in distributed and scalable IoT environments, providing flexible and fine-grained control over each access request \cite{ref7}.

 The policies and attributes can be adjusted to meet the specific requirements of each organization (such as adding or deleting certain attributes) \cite{ref7}. Additionally, ABAC enables access delegation and revocation to be carried out through simply adjusting the attribute values of the subject or object in order to grant or revoke access rights \cite{ref7, ref28}. Consequently, ABAC is well suited for dynamic and scalable IoT environments \cite{ref28}, as explained in more detail below.

\subsubsection{Attribute-Based Access Control (ABAC)}

We can define the ABAC as a logical access control mechanism that evaluates a set of attributes against predefined policies to determine the access decision (i.e., allow or deny). Attributes are set of properties that are defined and assigned by the admin to describe some features related to the following components: \cite{ref7, ref29}:

\begin{itemize}

\item Subject: It is the user or requester who initiates the request for access   to a specific resource or information asset. Subjects are characterized by their attributes. Subject's attributes include some identifying criteria such as user identity, role, organization, department, etc.

\item Object: It is the resource or information asset to which the subject is attempting to gain access such as an IoT device or data collected by an IoT device. The object's attributes may include identity, location, type, and owner.

\item Permission: the action or operation that the subject wants to apply to the object, such as read, write, or update. 

\item Environment: It represents the contextual information about the environment surrounding the access request when initiated, such as current time and place of the requester, as well as their client type (e.g. application, server, human, etc.).

\end{itemize}

All of the above attributes are assessed against pre-defined policies to decide whether or not to grant or deny access to the requester. The policy is a combination of the attributes related to the subject, object, permission, and environment; it outlines the access rules for what operations the subject is allowed to carry out over an object in specific environmental conditions.

\begin{figure}[htp!]
    \centering
    \includegraphics[scale=.45] {    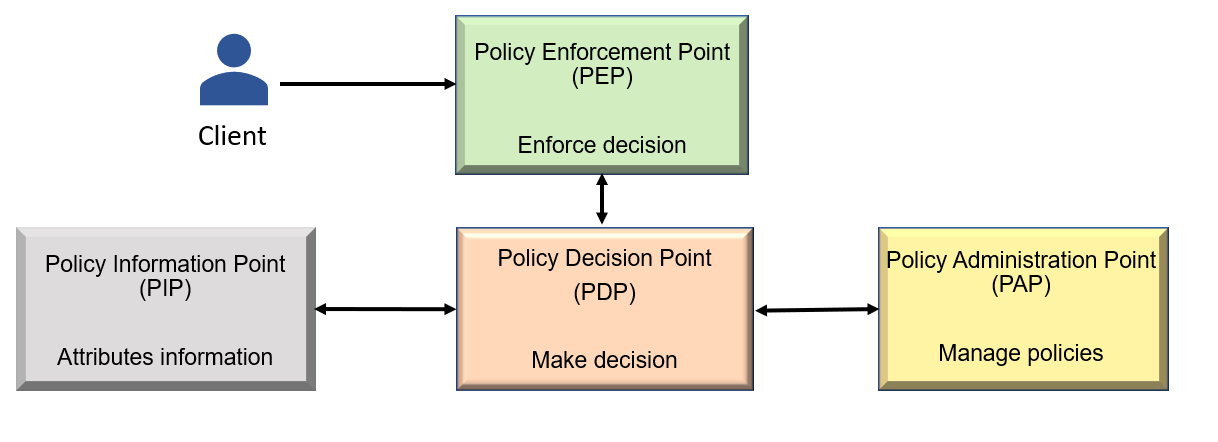}
    \caption{ABAC functional nodes}
    \label{fig1:ABAC functional nodes}
\end{figure}

As shown in Figure ~\ref{fig1:ABAC functional nodes}, ABAC contains four primary functional nodes that work together to carry out the access control process \cite{ref30, ref31}. These nodes are:

\begin{itemize}

\item Policy Enforcement Point (PEP): The PEP is responsible for transmitting the client's access request to the Policy Decision Point (PDP) to gain a decision on whether or not to allow access, and for enforcing that decision onto the client.

\item Policy Decision Point (PDP): It is the core of the system, responsible for providing access control decisions (allow or deny) to the PEP after assessing access requests against pre-configured policies and attributes from the Policy Administration Point (PAP) and the Policy Information Point (PIP), respectively.

\item Policy Information Point (PIP): It functions as a repository for the attribute values which the PDP needs in order to make access decisions. The PDP links up with the PIP to retrieve attributes of the subject, object, permission and environment required for policy assessment.

\item Policy Administration Point (PAP): It works as a repository for the policies and allows for policy management operations (such as adding, editing and deleting).

\end{itemize}

ABAC is centralized in terms of policy and attribute storage and management. Due to this reason, we aim in our research to decentralize the ABAC nodes (i.e., PDP, PAP and PIP) by incorporating them into a smart contract on the Blockchain.

\subsubsection{Data Access Policy Model}

Drawing inspiration from \cite{ref7}, we define an attribute as a pairing of its name and corresponding value, denoted as follows:
\begin{center}
\( A =  \{name: value\} \) \\
\end{center}


 In our model, each ABAC component (i.e., subject, object, permission, and environment) possesses one or more attributes denoted by SA, OA, PA, and EA respectively.

\(SA\) (Subjects' attributes): denotes the attributes of the subject (user or requester), comprising three elements: \(UserID\) (user’s unique identifier), \(Role\) (user’s role), and \(DomainID\) (identity of the domain in which the user is registered).

\begin{center}
\( SA =  \{UserID, Role, DomainID\} \) \\
\end{center}

\(OA\) (Objects' attributes): denotes the attributes of the object (data record that maintains the data collected by IoT device and other entered data), consisting of four elements: \(DeviceID\) (unique identification of the IoT device that issued the data), \(DomainID\) (identity of the domain in which the IoT device and the data are situated), \(OwnerID\) taking a value of the \(UserID\) to identify the data or device owner, and \(DataType\) (the type of collected data such as blood pressure, temperature, humidity, etc). The local administrator and data owner can simply delegate access rights of a specific IoT device by altering the value of the \(OwnerID\) attribute.

\begin{center}
\( OA =  \{DeviceID, DomainID, OwnerID, DataType\} \) \\
\end{center}

\(PA\) (Permissions' attributes): denotes the permissible operations that the subject can apply over specific objects such as read and write operations. The value of \(read\) and \(write\) can be 1 or 0 which indicates permission or prohibition of access respectively. The default value of \(PA\) for both operations is 1. Local administrators and data owners can  revoke access rights by modifying the operation's value to 0.

\begin{center}
\( PA =  \{Read, Write\} \) \\
\end{center}

 \(EA\) (Environmental attributes): denotes the requisite access control attributes of the environment, comprising three elements: \(AllowedIP\) is employed to restrict system access from any IP located outside the global network; \(StartTime\) and \(EndTime\) denote the time frame in which permitted users can access the requested data.

\begin{center}
\( EA =  \{AllowedIP, StartTime, EndTime\} \) \\
\end{center}

 We formulate the data access policy structure taking into account the ABAC model and the attributes of the IoT context. The policy \(P\) is depicted by a combination of the four components: \(SA, OA, PA,\) and \(EA\) as follows: 

\begin{center}

\( P = \{SA, OA, PA, EA\} \) \\

\end{center}

In the subsequent section, we show how to configure and design the smart contract to disseminate the ABAC functional nodes (i.e., PDP, PIP, and PAP).

\subsection{Smart Contract Configuration}

The smart contract is the core of our model, responsible for managing and enforcing access control decisions automatically. In the proposed model, we suggest implementing two smart contracts: Policy Contract and Access contract. The Policy Contract includes methods that enable the local admin to manage ABAC policies while Access Contract is the core of our system which implements access control functions for data requesters over IoT data. We describe the functions of each smart contract in what follows.

\begin{enumerate}

    \item Policy Contract (PC):  comprises the following functions to manage ABAC policies, after the authentication of the local admin:
    
     \begin{itemize}
    
    \item ValidatePolicy(): verifies the legitimacy of the ABAC policy. A valid policy must comprise all predetermined four attributes with the correct data type for each attribute.
    
    \item AddPolicy (): adds a new ABAC policy to the blockchain. The policy contract calls this function after affirming the validity of the policy with ValidatePolicy().
    
    \item UpdatePolicy(): enables the local admin to modify the values of existing policies and overwrite the old ones in accordance with the current situation. As with AddPolicy(), the Policy Contract must call ValidatePolicy() before updating the desired policy.
    
    \item DeletePolicy(): eliminates a particular ABAC policy in two cases. One case materializes when the admin actively invokes this method to delete a policy. The other instance occurs when the DeletePolicy() method is triggered automatically if a policy has expired when its "EndTime" attribute is attained.
    
    \item QueryPolicy(): retrieves the ABAC policy from the blockchain based on the subject's or object's attributes. It returns empty if no policy is linked to the request. Otherwise, it returns one or more pertinent ABAC policies.
    \end{itemize}
    
    \item Access Contract (AC): contains the following access control methods to check the matching of the ABAC request with the ABAC policy. These methods are called after authenticating the data requester.  In addition, these methods also facilitate the delegation of access rights between users.
    
     \begin{itemize}
     
     \item GetAtts(): is employed to acquire the values of the attributes from the ABAC request transaction; it is useful for retrieving information regarding a subject and object and any other related information.
     
     \item CheckAccess(): checks the matching between the subject and object attributes that are included in the ABAC request by invoking the GetAtts() method and compares it with the ABAC policy returned by the QueryPolicy() function. If all attributes are found to be compliant with the policy, access will be granted; otherwise, it will be denied.

   \item DelegateAccess(): is utilized to transfer rights of access among network users by altering the value of the \(OwnerID\) attribute. It invokes GetAtts() to check if the transaction is issued by the data owner itself. Then, it  calls UpdatePolicy() method from the Policy Contract to add another value to the \(OwnerID\) attribute.
   \end{itemize}
    
\end{enumerate}

To sum up, the PC functions as a policy administration point (PAP), which is responsible for managing ABAC policies, while the AC acts as both a policy information point (PIP) and policy decision point (PDP) and is in charge of handling attributes and making access decisions, respectively.

\subsection{ABAC Components Placement }

 In this section, we illustrate how  the attribute-based access control functional nodes are distributed and combined within the proposed system architecture. 
 
\begin{figure}[htp!]
    \centering
    \includegraphics[scale=.31] {    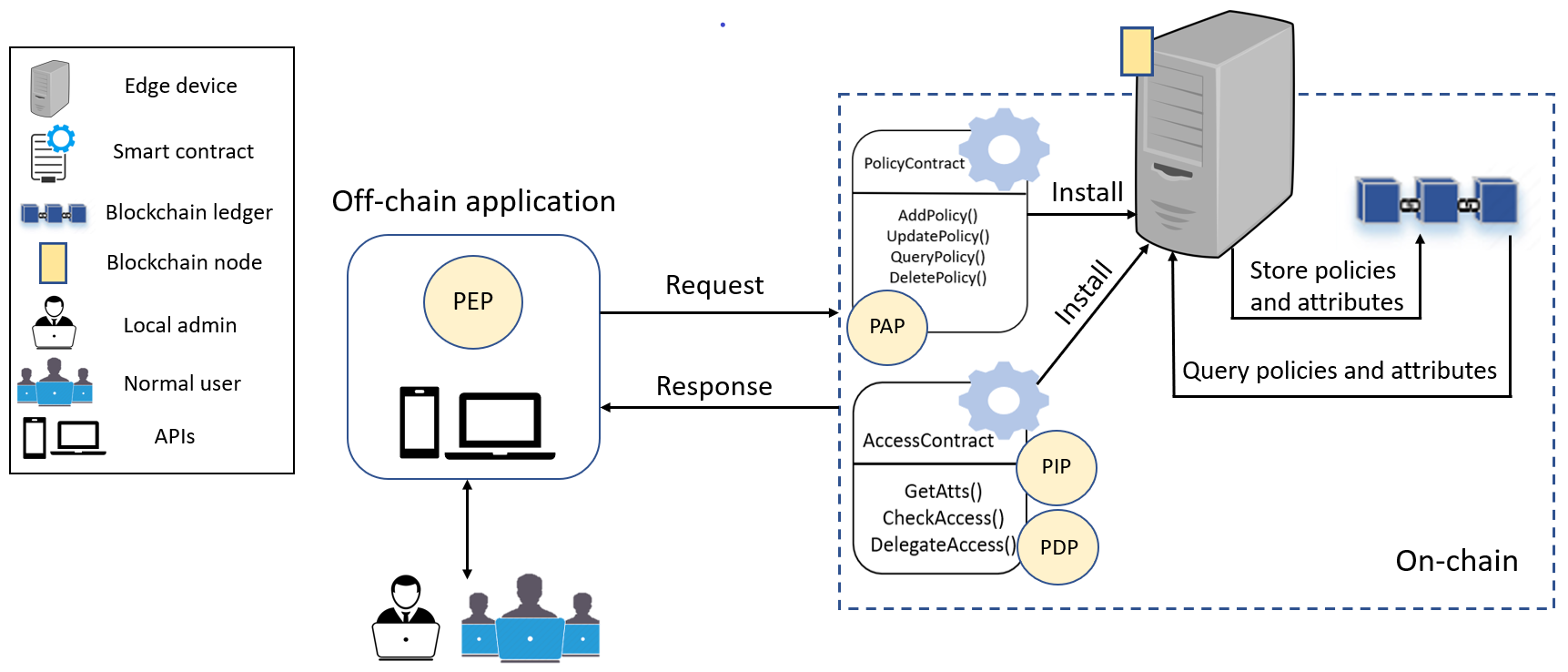}
    \caption{System architecture with ABAC}
    \label{fig1:ABAC components placement}
\end{figure}

As highlighted in Figure ~\ref{fig1:ABAC components placement}, the ABAC functional nodes are situated in two principal parts. The first part is an off-chain system that contains the client applications used to request access permissions from an on-chain part (e.g., blockchain system). The policy enforcement point (PEP) is situated within this off-chain area. The second part is an on-chain system that oversees ABAC components (i.e., PAP, PIP, and PDP) through smart contracts and securely stores the values of the attributes and policies in a tamper-proof ledger.

The user initiates the ABAC request through the client application (PEP), which in turn forwards the  request to the edge device in the on-chain part. The edge refers to the access contract to make the access decision  by invoking CheckAccess() method (PDP). CheckAccess() communicates with PIP and PAP by calling GetAtts() in AC and QueryPolicy() in PC to query the related attributes and policies, respectively, from the blockchain ledger. After receiving and assessing the required attributes and policies, the access contract makes the access decision and returns it to the client application to enforce the access decision.

\subsection {System Workflow}

 In this section, we elucidate the general system workflow, which is broken down into three primary phases. We additionally delineate the intermediary steps of each phase illustrated in Figure~\ref{fig1:workflow1}.

 \begin{itemize}
     \item Phase 1: Initialize the blockchain network and deploy the smart contract. These operations are implemented by the global admin.
     
    \subitem Step 1: Set up the blockchain network by initiating the identities (i.e., certificates and secret key pairs)  for all network entities (i.e., users and devices). The issued certificates encompass the attributes needed to restrict access.
    
    \subitem Step 2: Define the smart contract and deploy it on the blockchain network.
     
   \item Phase 2: Define and upload the access control policies into the blockchain network. 
     
     \subitem Step 1: Define access control policies. The local admin and the data owners need to work together to decide on the access policies based on the attributes of the subject, object, permission, and environment.  
     
     \subitem Step 2: After the access policy is defined, the local   admin uploads it to the blockchain.
     
      \subitem Step 3: By connecting to the blockchain network, the local admin can run Policy Contract to add, delete, and update policy. The data owner also can run the Access Contract to delegate access right over his own data record to another user. The value of the policy is saved in the on-chain storage. 
     
     \item Phase 3: The process of verifying the access right (authorization) of the data requester  is fundamental to our system. We have two scenarios based on the position of the data requester and the requested data:
     
     \subitem Step 1: The data requester sends an ABAC request based on the attributes of the subject, object, and permission.
     
     \subitem Step 2: Invoke the Access Contract to execute the CheckAccess() method (See section 4.5),   apply attribute-based access control, and   get the access decision (i.e., approve or reject).

       \subitem Step 3: If  the access decision results in  a reject, the edge node returns an error message. On the other hand, if it returns approve, the edge node assesses the location of the requested data by equating the DomainID attribute for both  the subject and object for data retrieval. We have two cases  to extract data when the access decision returns approve depicted in Figure~\ref{fig1:workflow2}:

      \begin{figure}[htp!]
    \centering
    \includegraphics[scale=.35]{    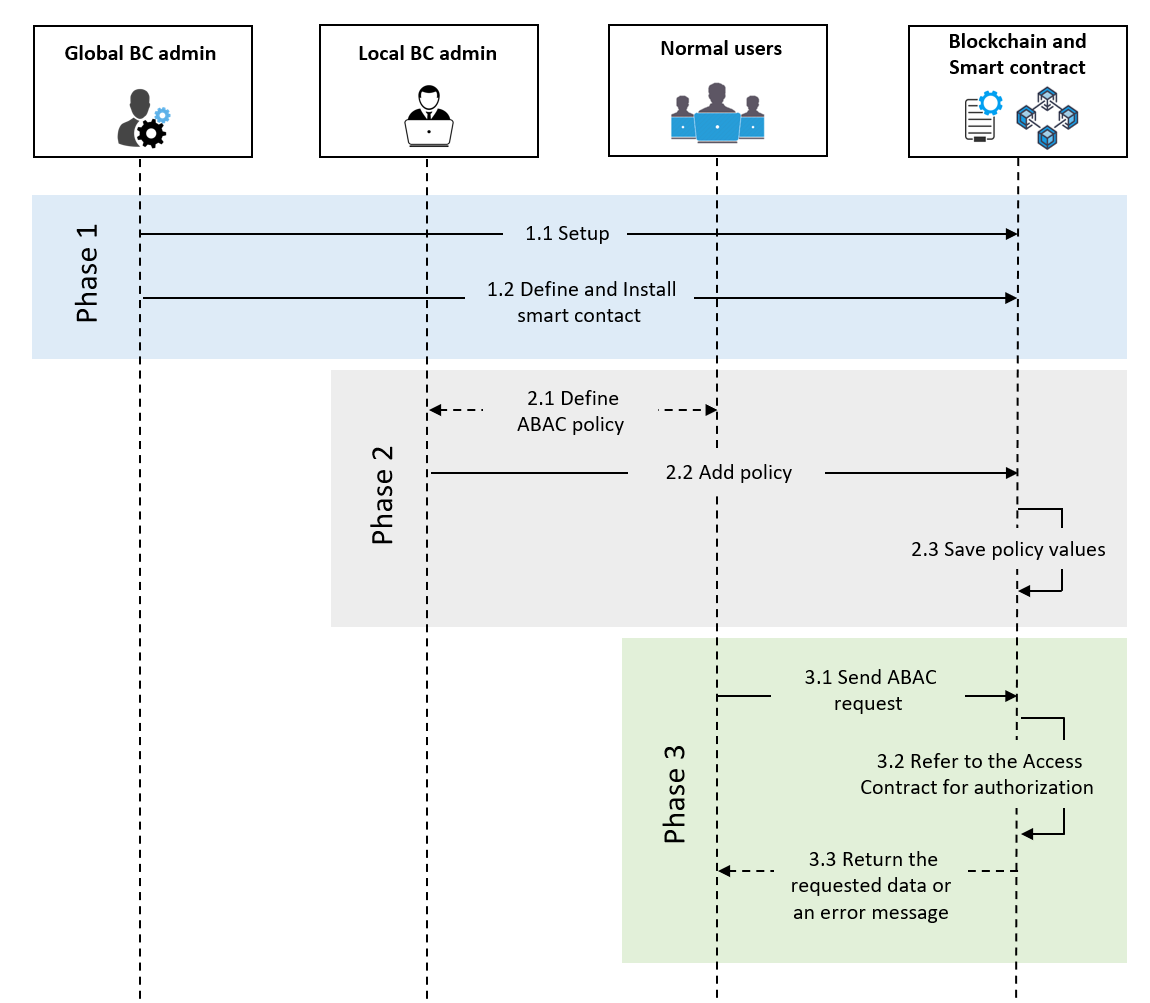}
    \caption{System workflow }
    \label{fig1:workflow1}
\end{figure}

      \begin{figure}[htp!]
    \centering
    \includegraphics[scale=.37]{    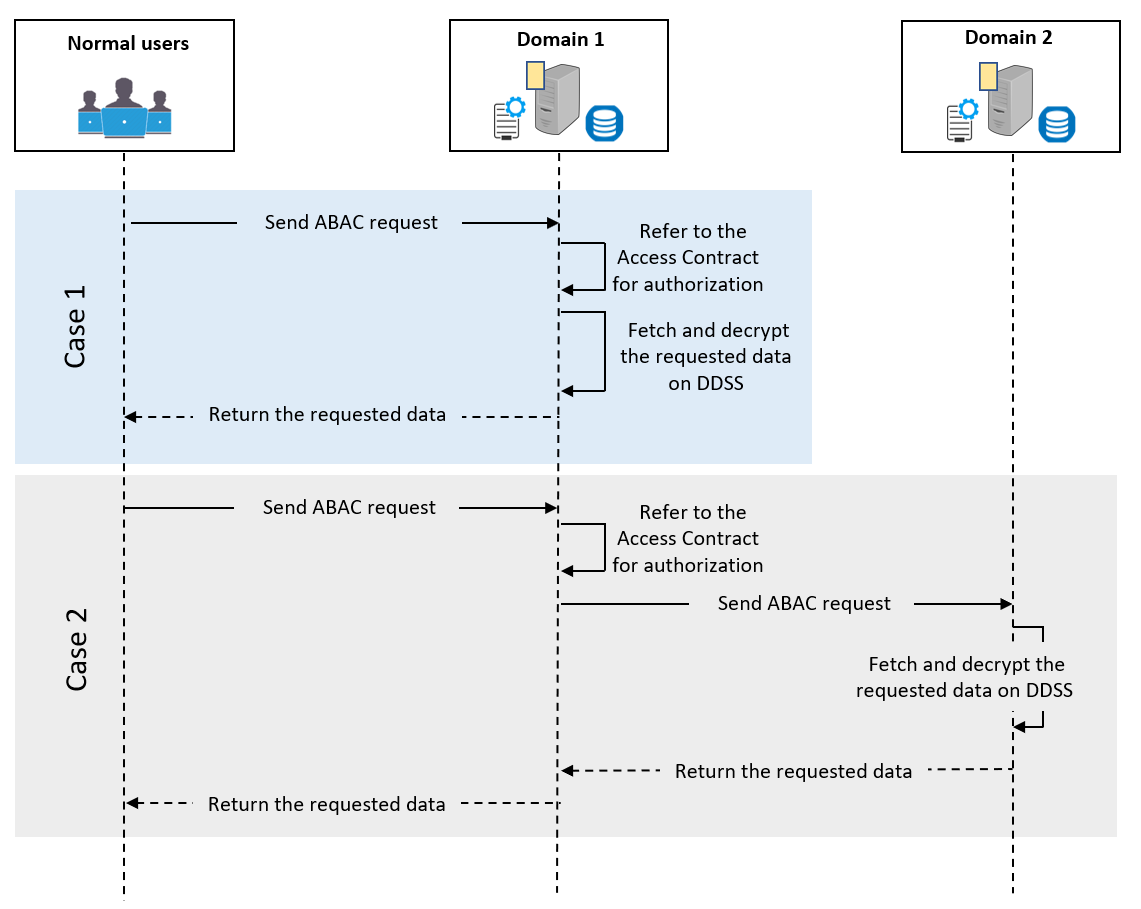}
    \caption{Phase 3 data retrieval workflow }
    \label{fig1:workflow2}
\end{figure} 
       
       \subsubitem Case 1: When the data requester and requested data are located in the same domain.
       
       \begin{itemize}
         
      \item If matching = true, the edge fetches and decrypts the requested data directly on DDSS.
      
      \item  Return the requested data to the data requester.
       
         \end{itemize}
       
       \subsubitem Case 2: When the data requester and requested data are located in different domains.
       
      \begin{itemize}

      \item If matching = false, the edge identifies the destination domain
      that stores the requested data.
  
      \item The edge forwards the access request to the destination edge node for data retrieval.
      
      \item	The destination edge fetches and decrypts the requested data on DDSS.
      
      \item It returns the requested data to the previous edge that then passes it to the data requester.

      \end{itemize}
     
 \end{itemize}

\section{Performance Analysis}\label{sec:test}

In this section, we present the proof-of-concept implementation in section 4.1 and the testing phases in section 4.2.

\subsection{Implementation}

We introduce here a detailed description of our work. Section 4.1.1 provides an overview of the Hyperledger Fabric blockchain platform along with the key design features. Section 4.1.2 lists and describes the key concepts of the Fabric network.

\subsubsection{Hyperledger Fabric (HLF) }

To build our blockchain-based access control model and to control access to IoT data, we choose the Hyperledger Fabric Blockchain  (HLF) platform. HLF is an open-source blockchain platform launched by the Linux Foundation in contribution with IBM in July 2017 \cite{ref28, ref7}. Some design features of the Hyperledger Fabric make it a suitable platform for most IoT use cases. One of these features is that HLF is a permissioned blockchain platform in which all participants must be authorized before joining the blockchain network. Another important feature is the modularity structure that helps the Fabric meet a broad range of business use cases and enterprise purposes (i.e., healthcare, banking, smart cities, supply chain, government and so much more). The modular architecture of the Fabric supports pluggable components including smart contracts, consensus algorithms, storage systems, encryption algorithms and other services. Pluggable consensus algorithms used by Fabric including Solo, Kafka, and Raft require less processing and electric power to execute transactions compared to PoW and PoS and eliminate the need for costly mining operations. According to that, transaction execution has no fees.

Unlike other blockchain platforms (either permissionless like Ethereum or permissioned such as Quorum)  that support the order-execute model to perform sequential transaction execution, Fabric introduces a new transaction architecture called execute-order-validate that allows parallel execution. This new architecture enhances the performance and scalability by separating the transaction life cycle into three phases \cite{ref10}:

\begin{enumerate}

    \item Execute: Check the correctness of the transaction for execution and endorsement in any order.  

    \item Order: Use the pluggable consensus algorithm to order the transactions and disseminate them to other peers.

    \item Validate: Ensure that the transactions satisfy the endorsement policies before committing them to the ledger.
    
\end{enumerate}

Indeed, the order-execute model enforces the blockchain platforms to write the smart contract using a domain-specific programming language such as Solidity to eliminate non-deterministic operations. On the contrary, the above new architecture allows Fabric to write the smart contract using basic programming languages including Java, Node.js, and Go. Thus, enterprises and individuals do not need to spend much effort and time learning new programming languages. Furthermore, the execute-order-validate architecture utilized by Fabric filters out any inconsistency resulting from the first phase before ordering the transactions.

HLF also produces another key feature that boosts the performance and scales the system which is using the endorsement policy. Unlike other blockchain platforms where all participating nodes are required to execute transactions, Fabric uses endorsement policies to identify a subset of nodes  (i.e., n out of m) called endorsing peers to execute and endorse transactions \cite{ref10}. This adds a level of privacy and confidentiality compared to a public, permissionless blockchain network. Moreover, to increase the level of privacy and confidentiality, Fabric enables privacy and confidentiality through its private data feature and channel architecture. \cite{ref40}

\subsubsection{Hyperledger Fabric Key Concepts}

In this section, we list and describe the key concepts of the Hyperledger Fabric design and structure  as follows:

\begin{enumerate}

\item Shared Ledger

Like other blockchain platforms, HLF has a ledger subsystem; however, the Fabric ledger differs by including two parts: the world state and the transactions log (blockchain), as shown in Figure ~\ref{fig1:fabricledger}. The word state is a ledger's database that stores the ledger records as key-value pairs, which reflect the current state of the ledger at a given time and is updated with each transaction. It is convenient to have the current value of an object readily available, as opposed to having to go through the entire blockchain to calculate it. The world state supports two types of state databases: CouchDB and LevelDB. As shown in Fig~\ref{fig1:CouchDB}, we choose to implement CouchDB. The transaction log or blockchain is the second part of the Fabric ledger. It is a distributed, immutable, and append-only log that records all transactions that appear within the network. The transaction log keeps track of all the changes made to the world state, providing a history for the world state. 

   \begin{figure}[htp!]
    \centering
    \includegraphics[width=0.6\textwidth]{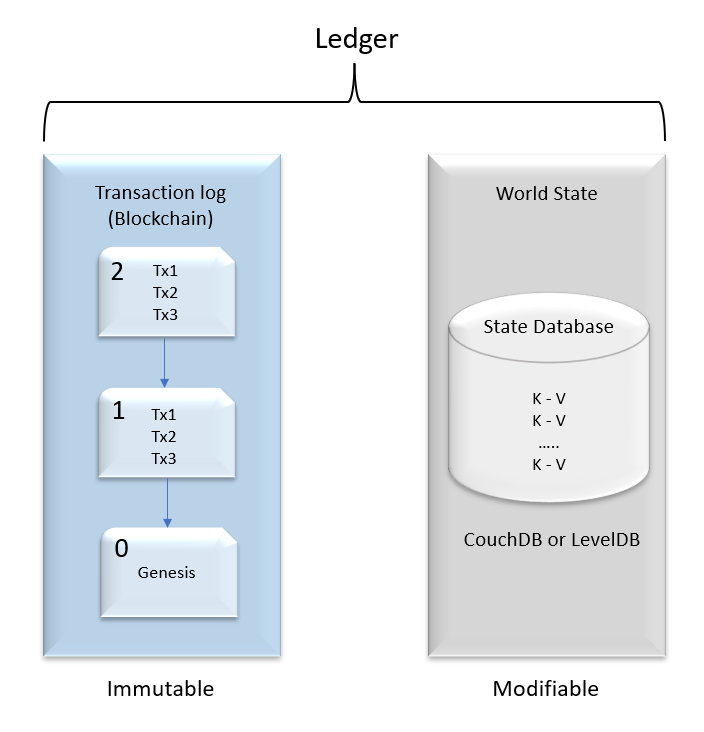}
    \caption{Structure of the ledger in Hyperledger Fabric }
    \label{fig1:fabricledger}
\end{figure} 

 \begin{figure}[htp!]
    \centering
    \includegraphics[scale=.29]{  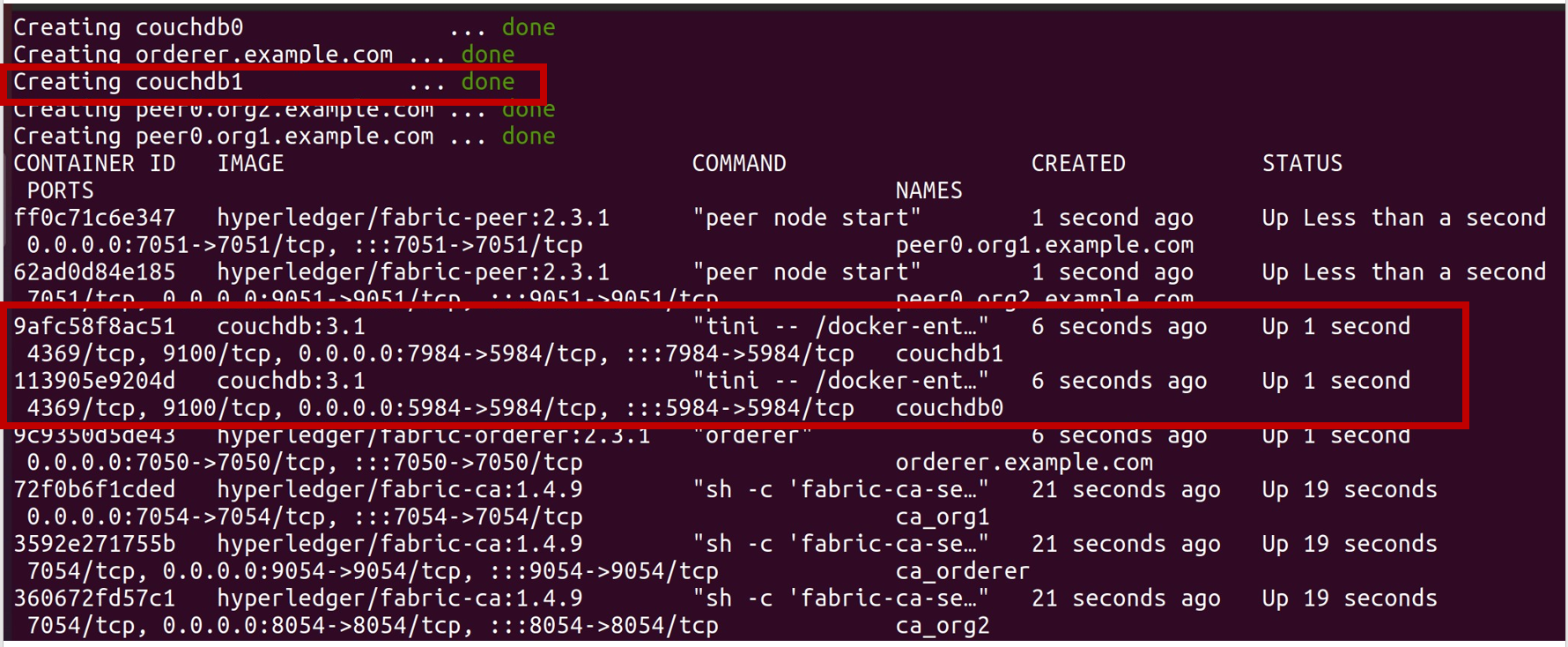}
    \caption{CouchDB as a state database }
    \label{fig1:CouchDB}
\end{figure} 

\item Peer

 Peer nodes (or simply peers) are the basic components of the HLF network, as they are responsible for hosting ledgers and executing chaincode (i.e., smart contracts). Each peer node maintains a copy of the shared ledger to ensure data integrity and consistency across the network. Furthermore, each peer can store multiple ledgers, each of which can be managed by one or more chaincodes.  Peers can be divided into two categories: endorser and committer. The endorser peer is responsible for executing chaincode and verifying the validity of a proposed transaction (i.e., endorsement). If the transaction is valid, it will be signed by the endorser and returned to the client. After verifying that all endorsements from endorsers are valid, the committer peer will save (i.e., commit) the endorsed transaction to the ledger, updating both the blockchain and ledger status. However, a peer node can also act as both an endorser and a committer at the same time. From the proposed architecture in the previous chapter, we choose the edge devices in each domain to act as peer nodes. Two peers are added to our network as shown in Fig~\ref{fig1:peer}.

  \begin{figure}[htp!]
    \centering
    \includegraphics[scale=.28]{  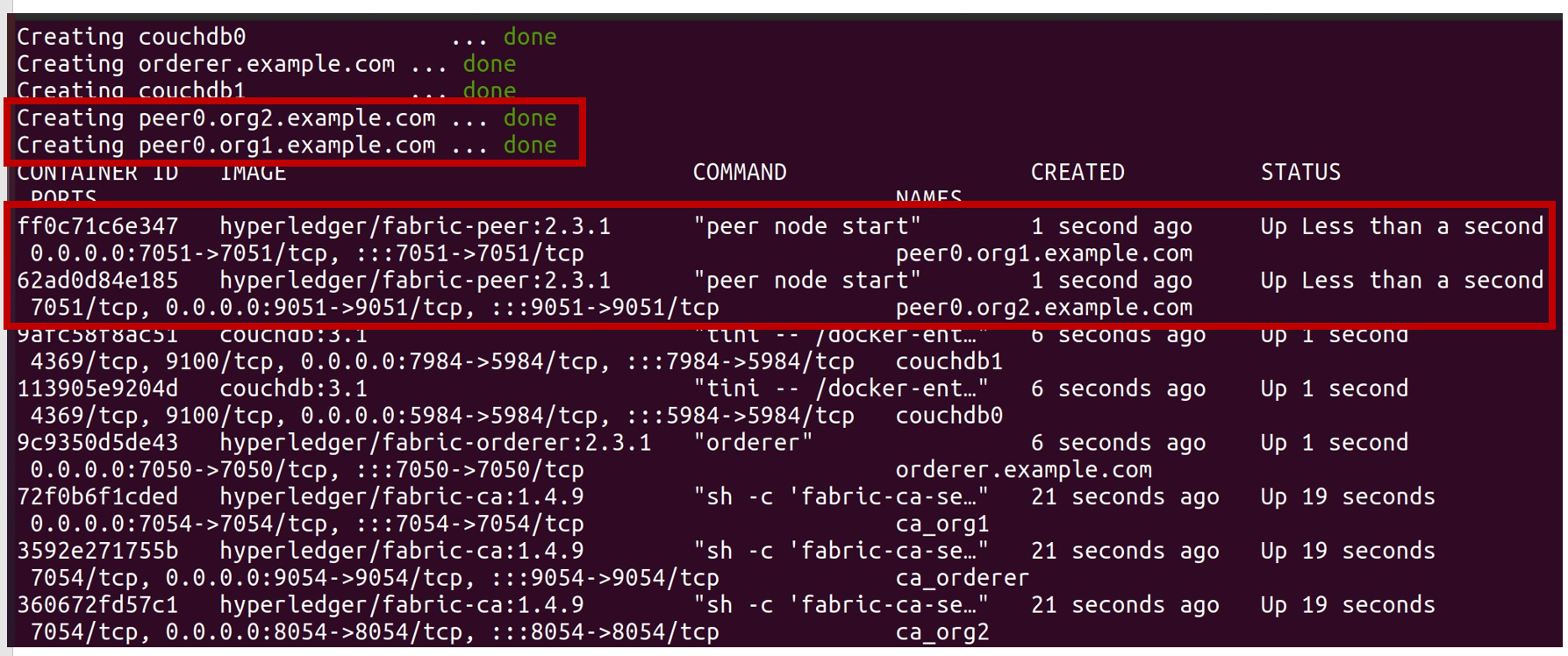}
    \caption{Peers creation }
    \label{fig1:peer}
\end{figure} 

\item Orderer

 Orderer (also known as an "ordering node") is a special type of Fabric node responsible for arranging transactions into blocks and then distributing them to all peers in the network. It also maintains consensus across the network by ensuring that all peers have the same view of the ledger. This differs Fabric from other blockchain platforms, such as Ethereum and Bitcoin, because it has a designated ordering node that is responsible for ordering transactions, rather than having all nodes do the ordering.  This improves the performance and scalability of the Fabric by preventing bottlenecks that can occur when all nodes are responsible for both executing and ordering tasks. There are three types of orderer in Fabric:

 \begin{enumerate}
 
\item Solo: it is a single ordering node used only for test purposes. The network implemented with Solo is not decentralized, has no fault tolerance, and cannot scale up with transactions. Solo is no longer supported and may be removed completely in future versions.

\item Kafka: Apache Kafka is a cluster of nodes implemented to provide crash fault-tolerant (CFT) ordering service. It utilizes a "leader and follower" model where each channel has a leader node responsible for making a decision which is then replicated by the followers. Kafka may require additional management overhead which makes it undesirable for many users.

\item Raft: like Kafka, Raft is a distributed consensus algorithm that allows a cluster of nodes to utilize the "leader and follower" model. It also supports CFT ordering services. The setup and management operations are easier in Raft than in the Kafka-based ordering service which makes it more recommended.

In our work, we decided to use the Raft ordering service (see Figure ~\ref{fig1:raftorderer} and Figure ~\ref{fig1:orderer} ).

 \end{enumerate}

 \begin{figure}[htp!]
    \centering
    \includegraphics[scale=.28]{  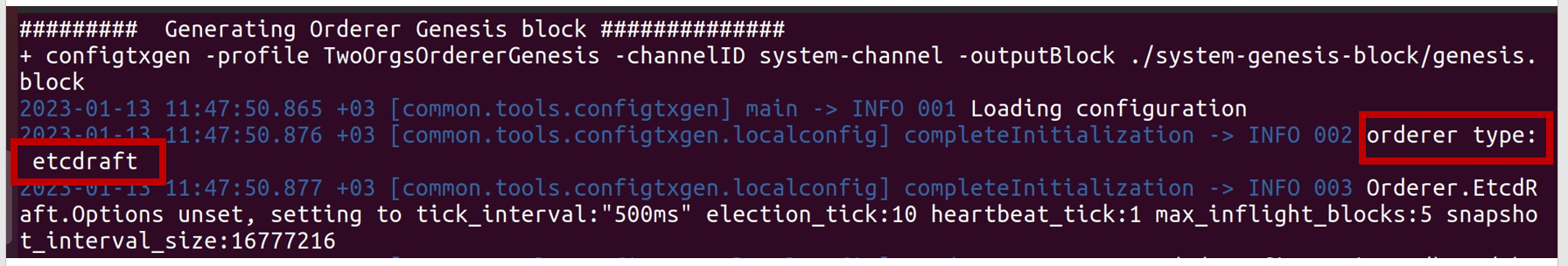}
    \caption{Raft orderer service}
    \label{fig1:raftorderer}
\end{figure}

 \begin{figure}[htp!]
    \centering
    \includegraphics[scale=.28]{  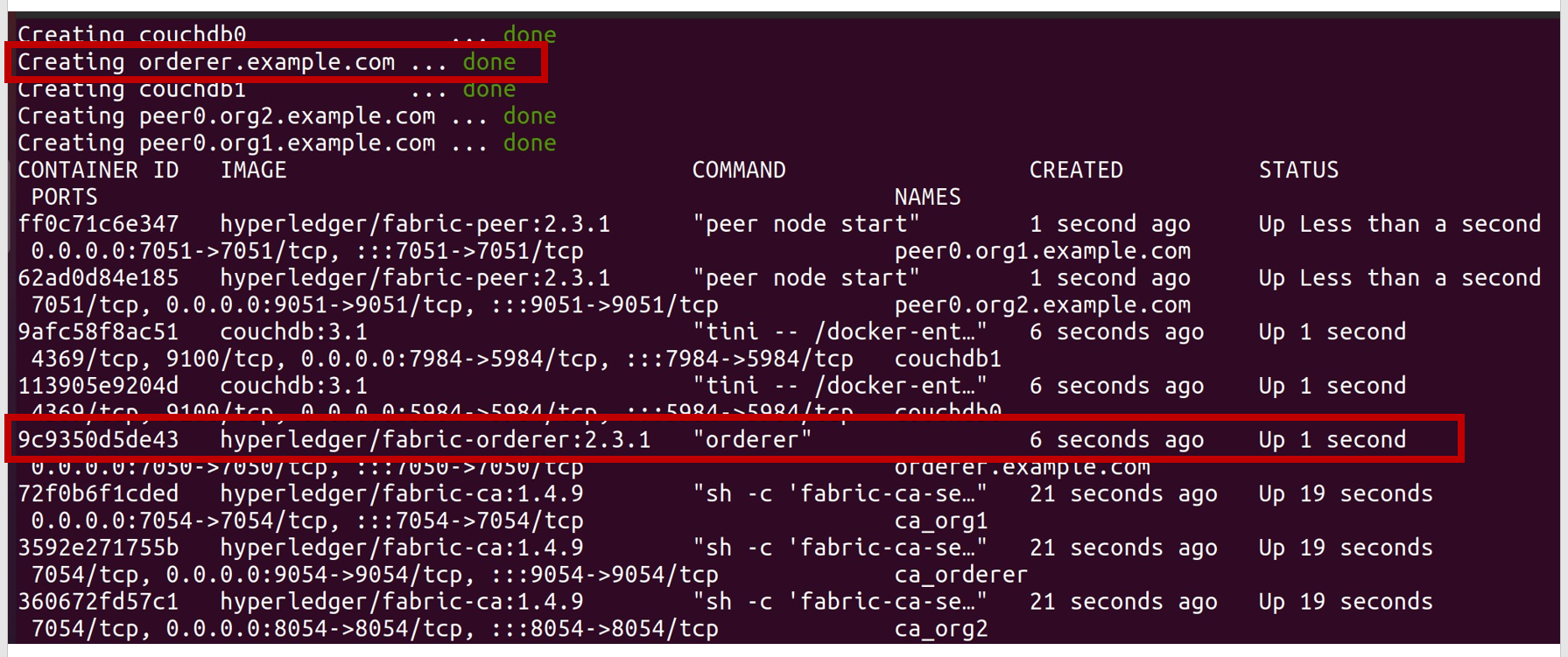}
    \caption{Orderer node creation}
    \label{fig1:orderer}
\end{figure} 

\item Client

Clint is basically an application created by SDK or a command line client that operates and manages the blockchain and interacts with peer nodes by sending transactions to the chaincode. The management operations applied by the client can be classified into two types. The first type is managing blockchain nodes by configuring, starting, or stopping nodes. The second type is managing the chaincode life cycle by installing, instantiating, upgrading
and executing the chaincode \cite{ref7}. In our work, we use the command line client to operate and manage the Fabric network and a third-party REST client called Postman to interact with the peer nodes by invoking chaincode.

\item Channel

The Fabric channel provides confidentiality and privacy for transactions by isolating "subnetting" communications between two or more organizations. Channel isolates transactions between different organizations because each channel has an independent shared ledger and chaincode applications. This feature allows business companies to hide their data from competitors. Figure ~\ref{fig1:channel} illustrates the creation of our channel with the name mychannel.

\begin{figure}[htp!]
    \centering
    \includegraphics[scale=.18]{  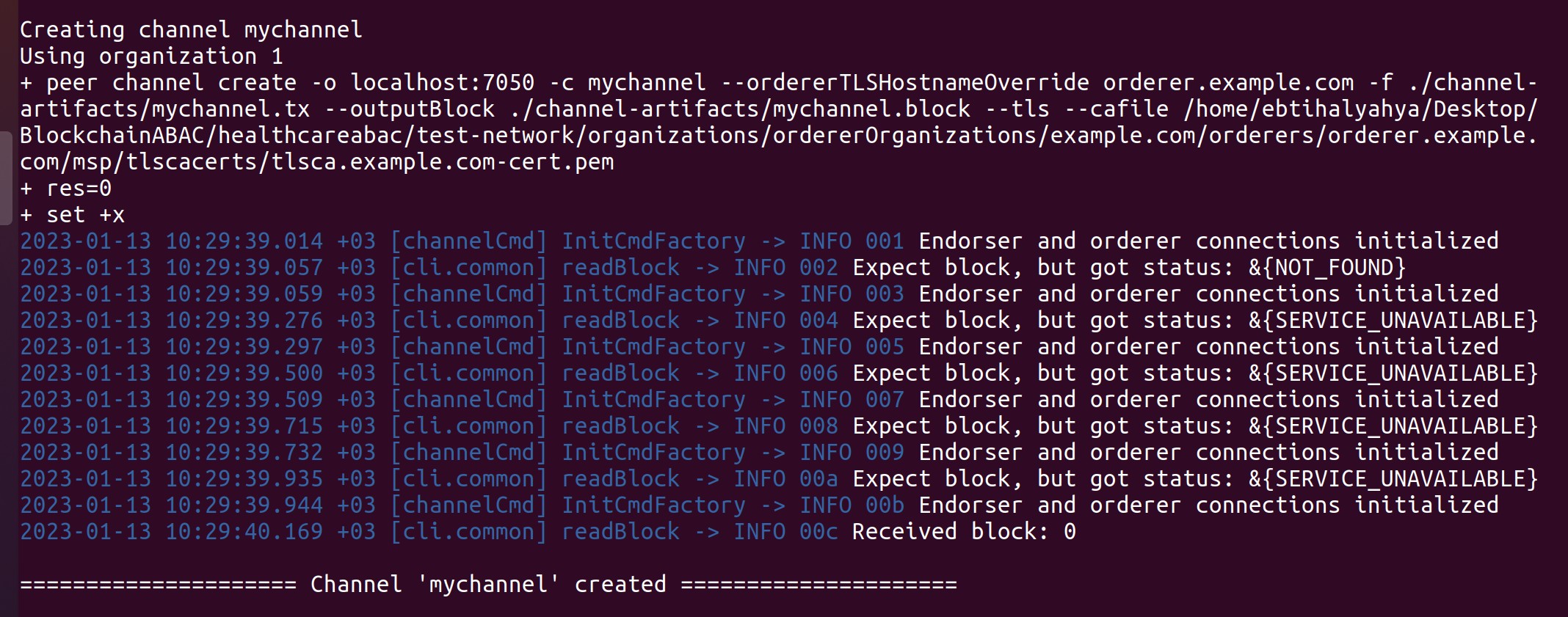}
    \caption{Channel creation}
    \label{fig1:channel}
\end{figure}

 \item Chaincode
 
Chaincode can be considered as a "smart contract" of the Fabric blockchain platform. It is a program or business logic running on the peer node to generate transactions. The generated transactions can read or change the data stored in the ledger state database (SDB). It is the only way for outsiders to communicate with the blockchain network. Developers may write multiple chaincodes to support multiple applications. Chaincode can be written in different programming languages such as Nodejs, Go, and Java.

\item Hyperledger Fabric CA

Fabric CA is the certificate authority for the Hyperledger Fabric blockchain responsible for issuing and registering  digital identities (i.e., certificates and secret key pairs) for network entities. Figure ~\ref{fig1:generateCert1} shows how we used Fabric CA to generate certificates for all blockchain members.

\begin{figure}[htp!]
   \centering
    \includegraphics[scale=.3]{  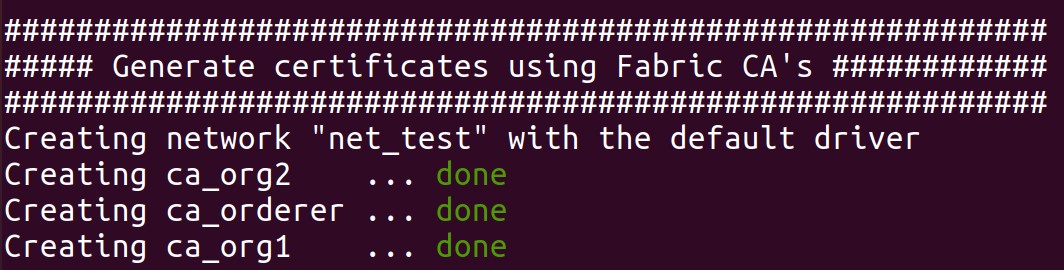}
   \caption{Certificates generation using Fabric CA's}
    \label{fig1:generateCert1}
\end{figure}

\end{enumerate}

\subsubsection{System Building Process}

As shown previously in chapter 4, the system workflow is divided into three phases. According to that, our experiment is implemented based on these phases.

\begin{itemize}

   \item Phase 1: Initialize the blockchain network and install the chaincode.

   \begin{enumerate}
    
    \item Generate identities (i.e., certificates and secret key pairs) for all network entities (i.e., channel, orderer, peer, admin and user) using Fabric CA. The issued certificates contain the required attributes to restrict access to a specific user. To do this step, We implement ABAC of Hyperledger Fabric as follows \cite{ref41}:

\begin{itemize}

\item Enroll global admin (registrar) in the CA to receive his certificate and the signing key (see Figure ~\ref{fig1:EnrollAdmin}).

\item Register user, peer, and domain admin by global admin into the CA. The CA then will return the secret (see Figure ~\ref{fig1:registeruser}).

\item Use secret to enroll user, peer, and domain admin to the CA. Then the CA will issue the certificates and signing keys for all these users. The issued certificates contain the needed attributes.

\item Generate Membership Service Provider Identity (MSPID) to the user, peer, and domain admin (see Figure ~\ref{fig1:GenerateMSP}).

 \end{itemize}

 As shown in ~\ref{fig1:generateCert}, the identities are generated for all network parties.

   \begin{figure}[htp!]
   \centering
    \includegraphics[scale=.15]{  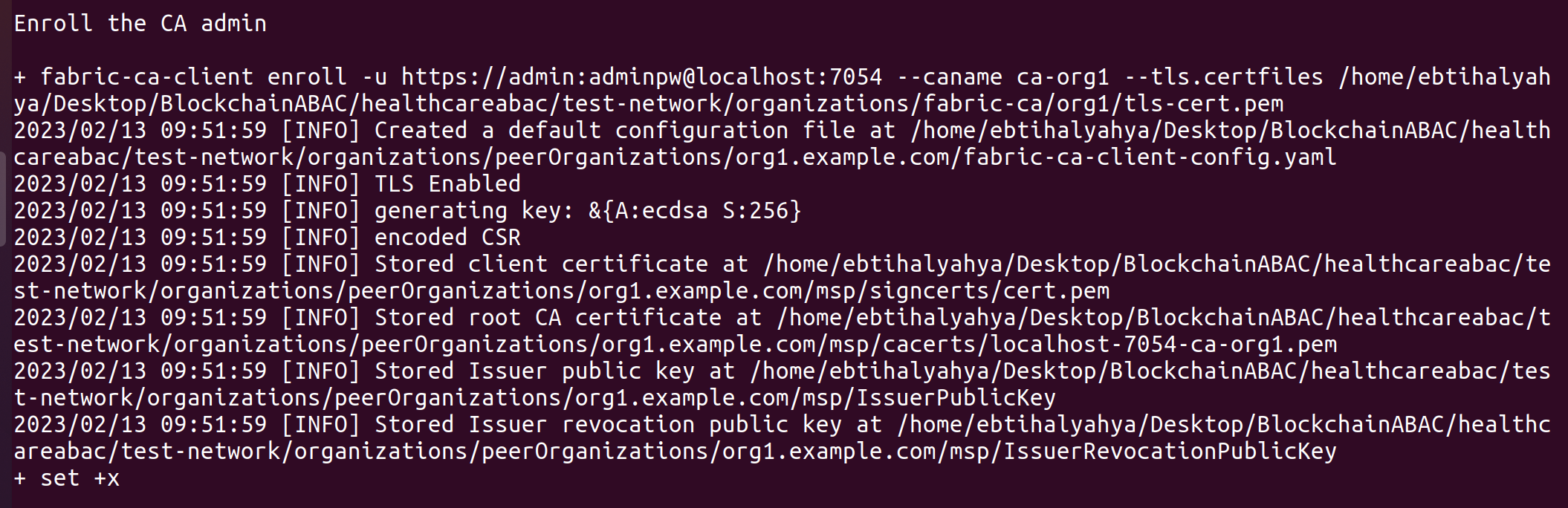}
    \caption{Enroll admin to the Fabric CA}
    \label{fig1:EnrollAdmin}
\end{figure}

   \begin{figure}[htp!]
   \centering
    \includegraphics[scale=.15]{  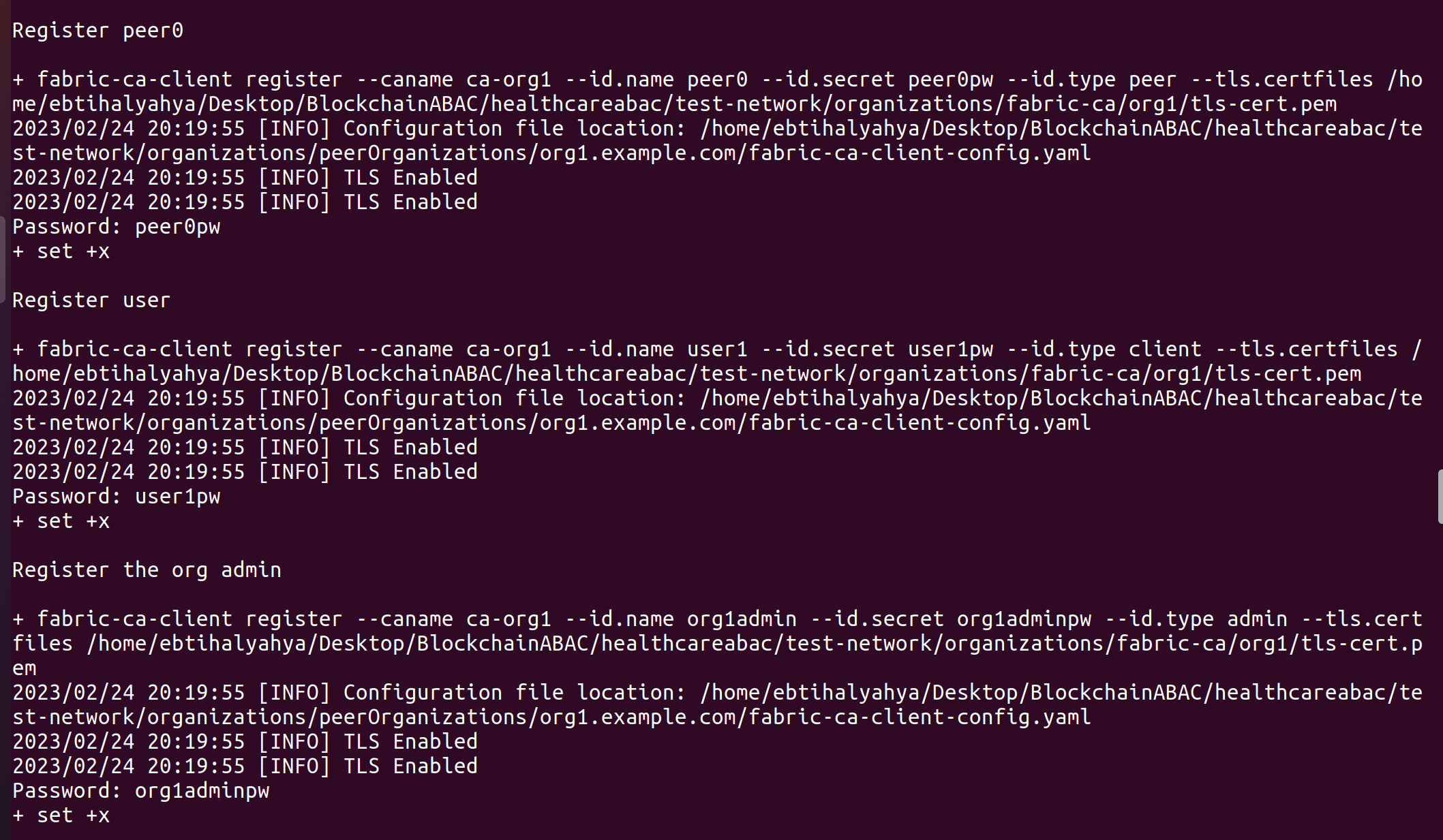}
    \caption{Register user to the Fabric CA}
    \label{fig1:registeruser}
\end{figure} 

\begin{figure}[htp!]
   \centering
    \includegraphics[scale=.15]{  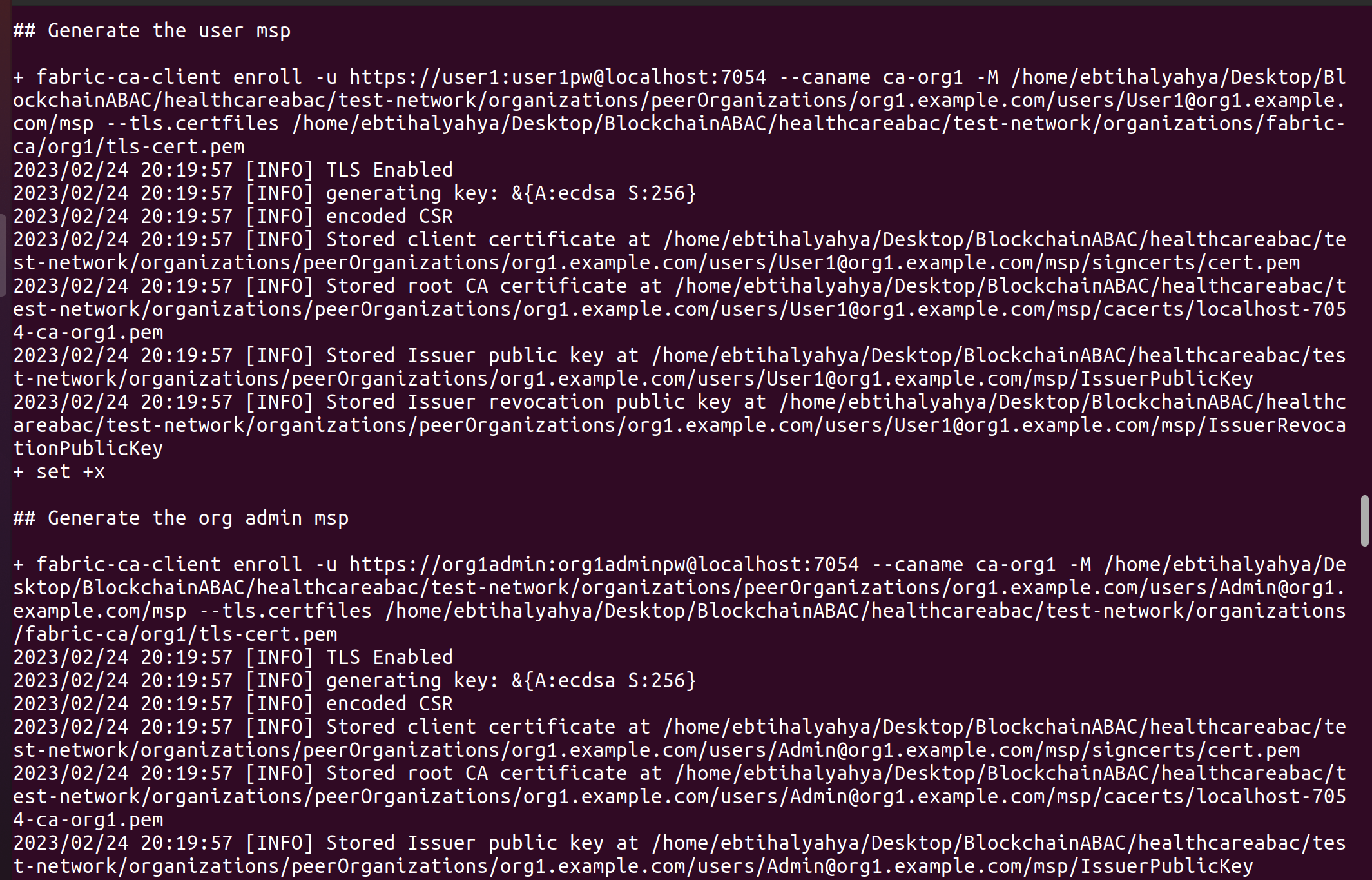}
    \caption{Generate MSP}
    \label{fig1:GenerateMSP}
\end{figure}

    \begin{figure}[htp!]
   \centering
    \includegraphics[scale=.25]{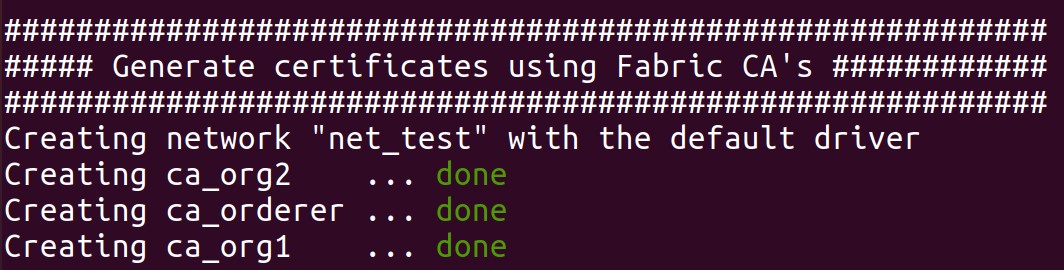}
    \caption{Certificates generation using Fabric CA's}
    \label{fig1:generateCert}
\end{figure}

\item Deploy the chaincode on the channel and install it to the peer nodes (see Figure ~\ref{fig1:DeployChaincode}).

 \begin{figure}[htp!]
   \centering
    \includegraphics[scale=.15]{  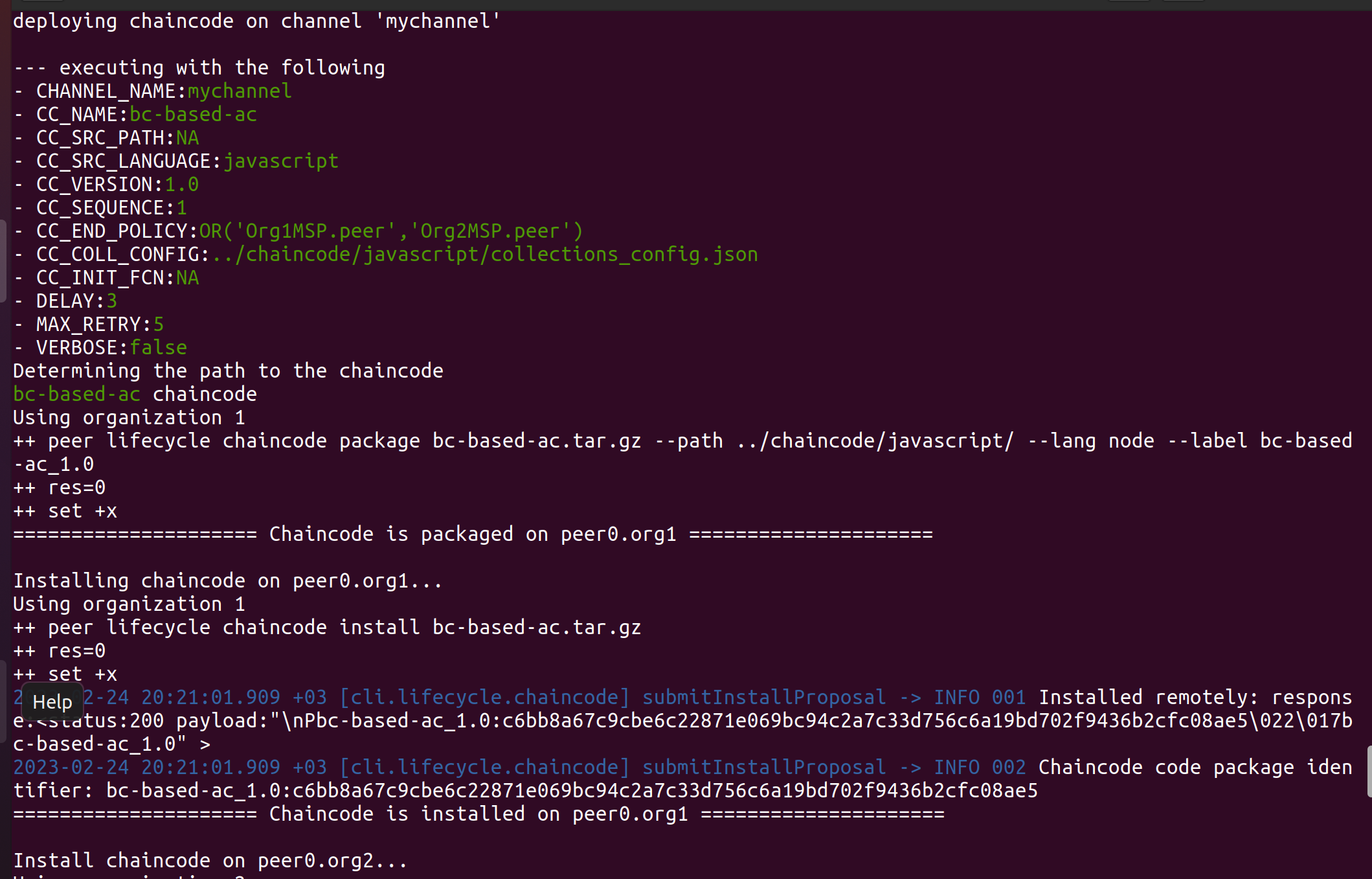}
 \caption{Chaincode deployment and installation}
    \label{fig1:DeployChaincode}
\end{figure}

 \end{enumerate}

\item Phase 2: Define and add the ABAC  policies into the blockchain
network.

\begin{enumerate}

    \item Define the ABAC policy in the chaincode based on the attributes of the subject (SA), object (OA), permission (PA), and environment (EA) as shown in Figure ~\ref{fig1:DefineABACPolicy}.

     \begin{figure}[htp!]
   \centering
    \includegraphics[width=0.4\textwidth]{  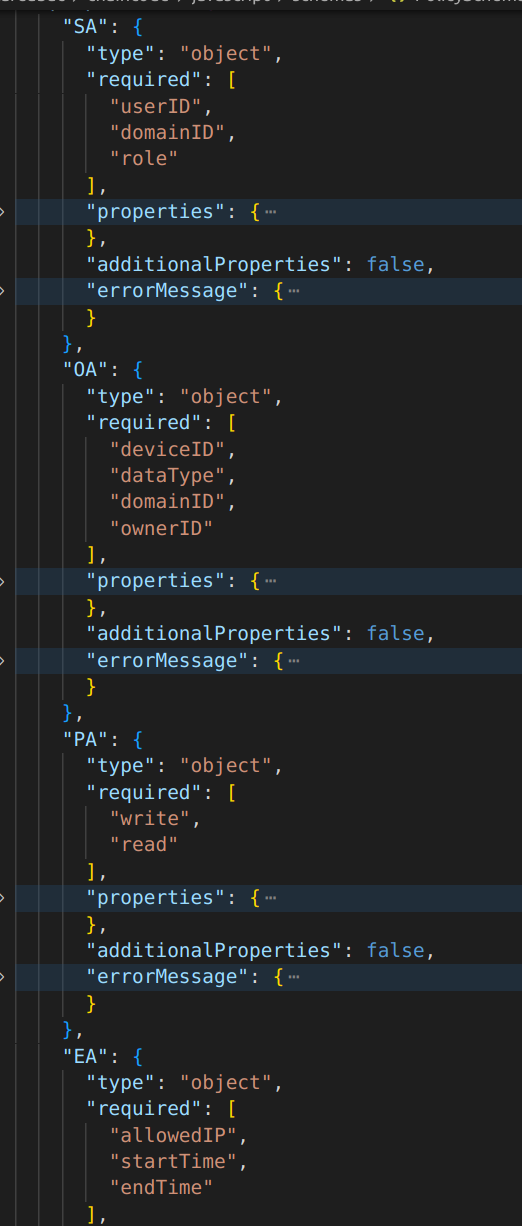}
 \caption{Example of ABAC policy definition }
    \label{fig1:DefineABACPolicy}
\end{figure} 

\item Add new ABAC policy to the blockchain  by invoking the AddPolicy() function of Policy Contract as shown in Figure ~\ref{fig1:AddPolicy}. The policy values are saved into CouchDB (i.e., Fabric state database).

\item By connecting to the Policy Contract, the local admin can query, update, and delete existing ABAC policy as follows:

\begin{itemize}

    \item Query existing policy by calling QueryPolicy() function to ensure whether is added successfully as shown in Figure ~\ref{fig1:QueryPolicy}.

    \item Update values of an existing policy by calling UpdatePolicy() method as shown in Figure ~\ref{fig1:UpdatePolicy}. The new  values override the old ones and the whole uttered policy is saved in the CouchDB.

   \item Delete existing policy from blockchain by calling DeletePolicy() function as shown in Figure ~\ref{fig1:DeletePolicy}.

\end{itemize}
   
\end{enumerate}

     \begin{figure}[htp!]
   \centering
    \includegraphics[scale=.28]{  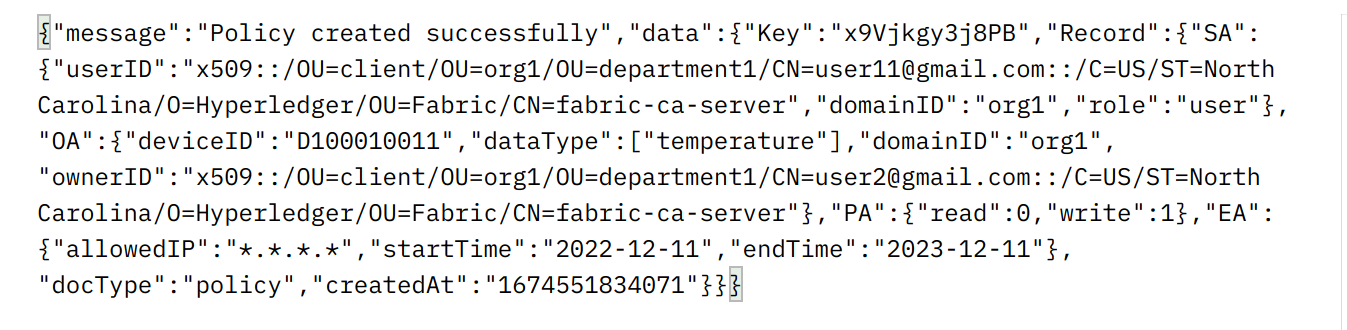}
 \caption{Result of adding new ABAC policy }
    \label{fig1:AddPolicy}
\end{figure}

   \begin{figure}[htp!]
   \centering
    \includegraphics[scale=.28]{  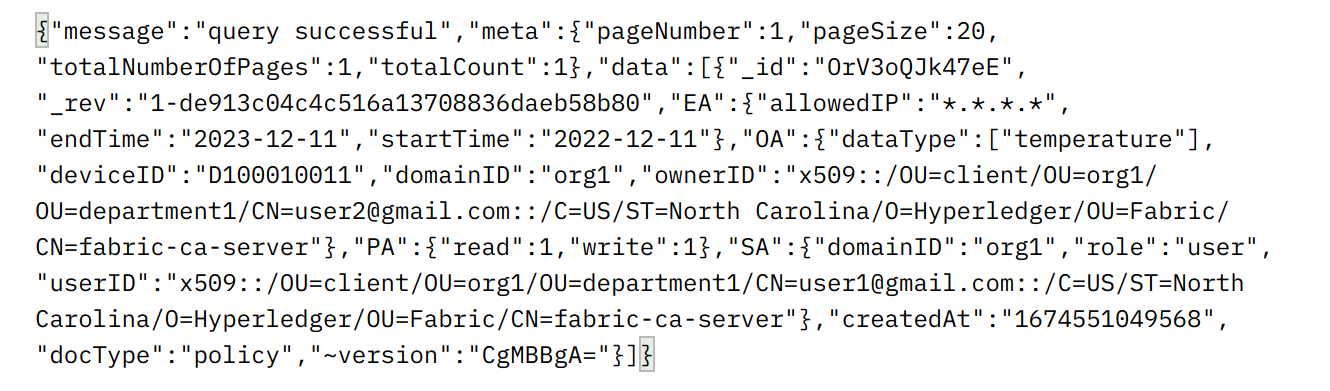}
 \caption{Result of querying existing ABAC policy }
    \label{fig1:QueryPolicy}
\end{figure} 

   \begin{figure}[htp!]
   \centering
    \includegraphics[scale=.28]{  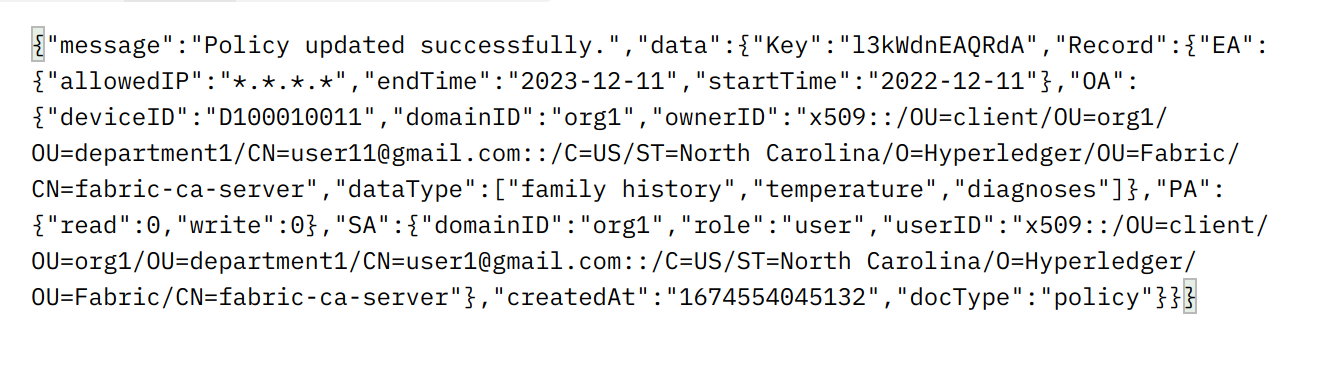}
 \caption{Result of updating existing ABAC policy }
    \label{fig1:UpdatePolicy}
\end{figure} 

  \begin{figure}[htp!]
   \centering
    \includegraphics[scale=.28]{  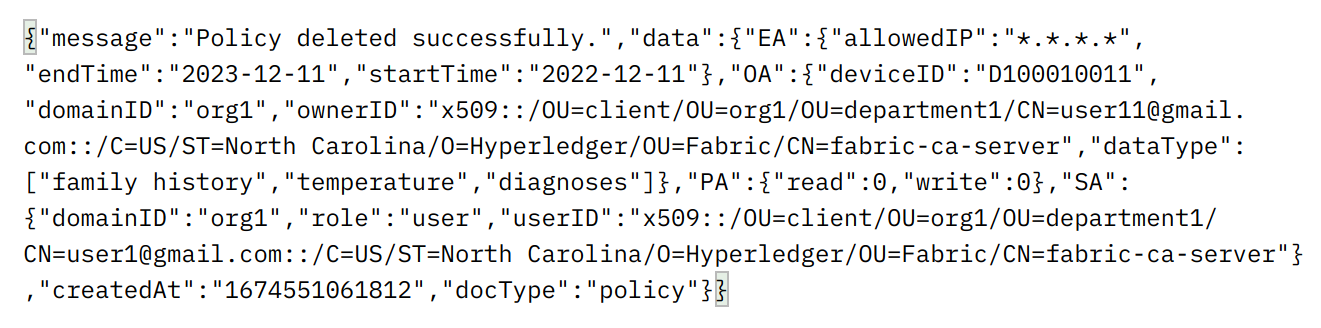}
 \caption{Result of deleting existing ABAC policy }
    \label{fig1:DeletePolicy}
\end{figure}

\item Phase 3: Check the access right (authorization) of the data
requester. In this phase, we only implement 
the access control part, which is applied locally within the domain that the data requester belongs to. We keep the data retrieval process for future work.

\begin{enumerate}

    \item Generate ABAC request by the data requester based on the attribute of the subject and object.

    \item Refer to the Access Contract to invoke CheckAccess() function to compare the receiving ABAC request with an existing policy. We use getAttributeValue() of the Fabric ABAC to retrieve the attribute value sent by the data requester and compare it with an existing policy by calling QueryPolicy() in Policy Contract.

    \item Figure ~\ref{fig1:CheckAccessApprove} and ~\ref{fig1:CheckAccessReject} show the results when the access decision  returns "approve" and "reject", respectively.

\end{enumerate}

\end{itemize}

     \begin{figure}[htp!]
   \centering
    \includegraphics[scale=.28]{  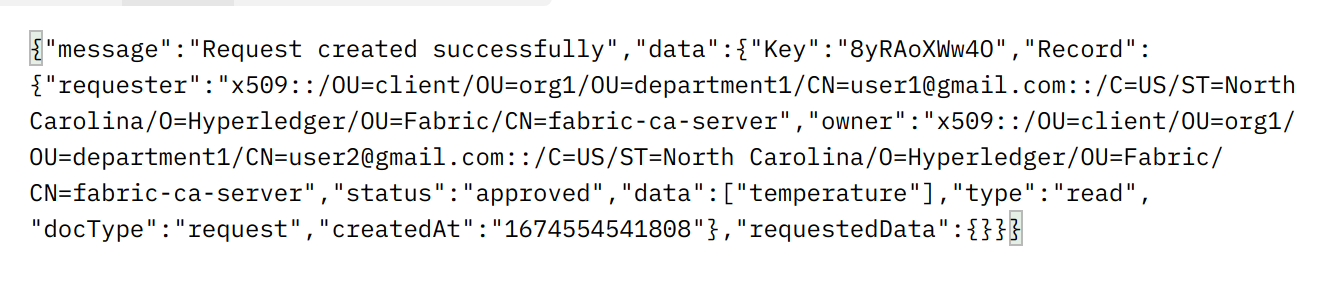}
 \caption{Result of check access when access decision returns "approve"  }
    \label{fig1:CheckAccessApprove}
\end{figure} 

 \begin{figure}[htp!]
   \centering
    \includegraphics[scale=.28]{  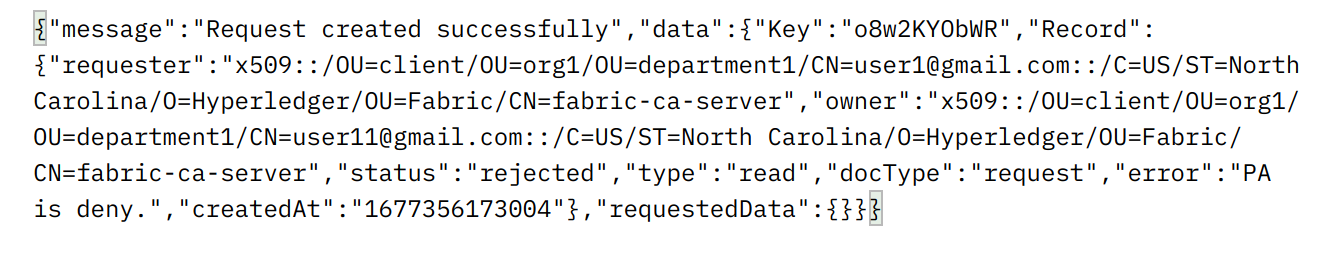}
 \caption{Result of check access when access decision returns "reject"  }
    \label{fig1:CheckAccessReject}
\end{figure} 
    
\subsection {Testing}

As discussed before, one of our objectives is to enhance the efficiency (i.e., reduce latency and increase throughput) of the distributed blockchain-based access control models. In this section, we provide a detailed description of the testing environment. Also, we measure and compare the average latency and throughput of the proposed works with the Fabric-IoT model \cite{ref7} and ABAC-HLFBC \cite{ref28}.

 \subsubsection{Testing Environment}

 The experiment of this work is carried out on a PC with an Intel(R) Core(TM) i7-7660U at 2.50GHz and 8.00 GB of RAM. It is implemented on Linux 64-bit Ubuntu 20.04 LTS installed on the local machine. The chaincode is written in Node.js and deployed on each peer node. A Third-party REST Client (Postman) is used to interact with the peer nodes used to invoke chaincode in the Hyperledger Fabric network. CouchDB is used to build the state database to store the policies and attributes' values.

\begin{figure}[htp!]
    \centering
    \includegraphics[scale=.5]{  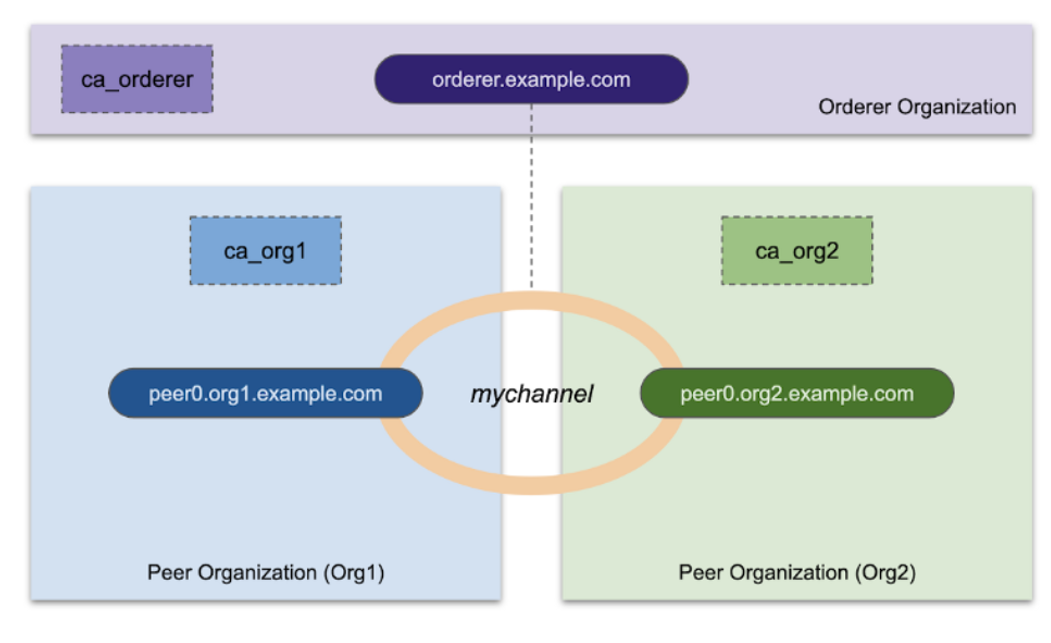}
    \caption{Network structure}
    \label{fig1:Network structure}
\end{figure}

 Figure ~\ref{fig1:Network structure} shows the prototype network structure that consists of one channel and three organizations: two peer organizations and one orderer organization. Org1 and Org2 represent the two peer organizations each with one peer node (i.e., peer0). The orderer organization contains a single orderer node implemented with Raft. Each organization has one certificate authority (CA).

\begin{figure}[htp!]
    \centering
    \includegraphics[scale=.2]{  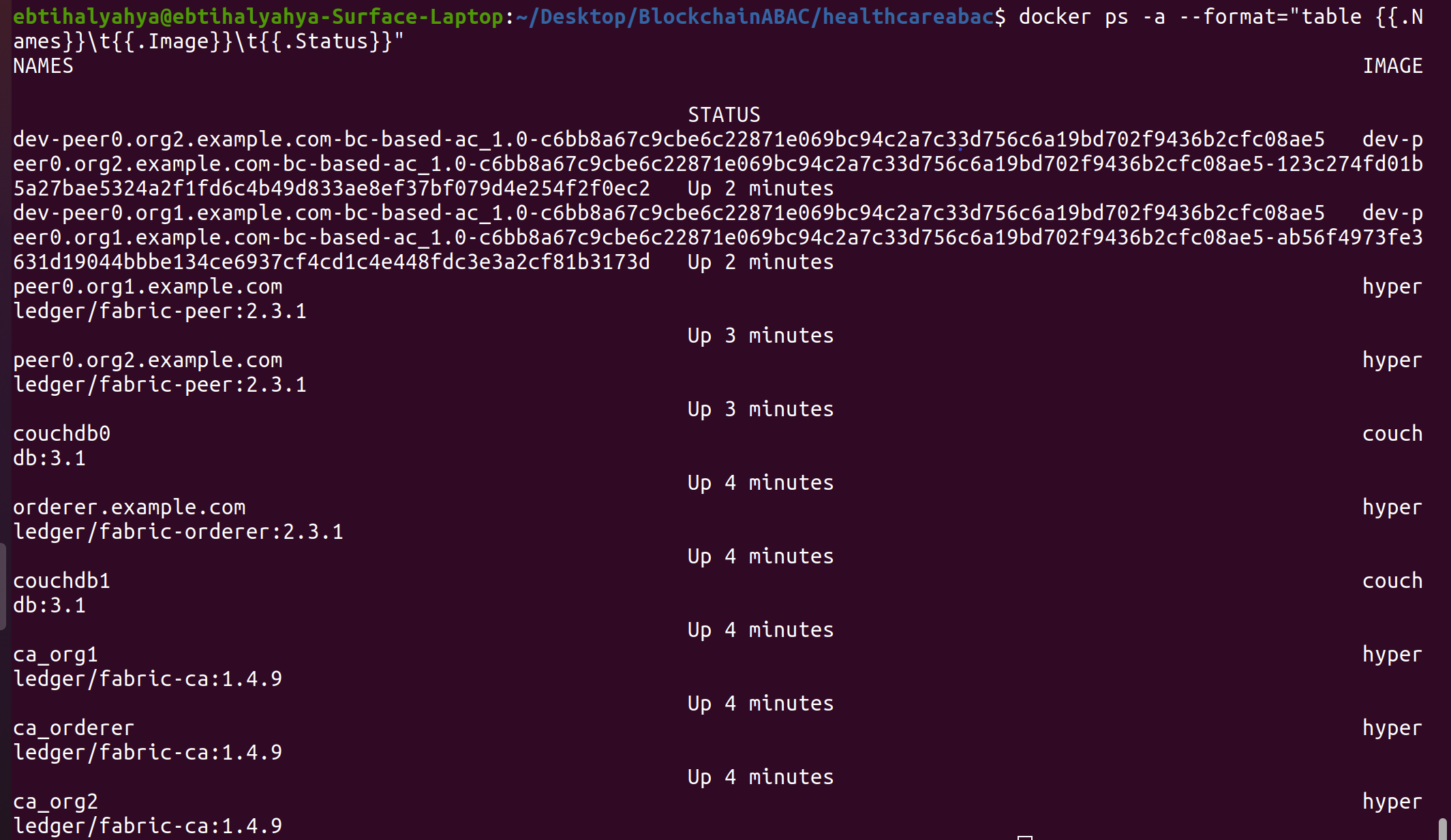}
    \caption{Docker container images}
    \label{fig1:Docker container image}
\end{figure}

Docker container is used to configure and manage the network entities. As shown in Figure ~\ref{fig1:Docker container image}, five types of Docker images nodes are used in our DBC-ABAC system for the following components:

 \begin{enumerate}
    
    \item Peer containers
    \item Orderer containers
    \item CouchDB containers
    \item Fabric CA containers
    \item Chaincode containers
    
\end{enumerate}

Table ~\ref {tab: Docker images of nodes} lists all Docker images of nodes  along with their type and number.

The endorsement policy "Peer0:org1 AND Peer0:org2" indicates both Peer0 located at org1 and org2 are required to sign transactions to ensure their validity.

In order to measure the performance of the proposed DBC-ABAC model, we use Hyperledger Caliper \cite{ref42}. It is a benchmark tool to test blockchain performance in terms of transaction latency and throughput and other performance metrics \cite{ref10}. Table ~\ref{tab:Hardware and software environments} lists the hardware and software requirements.

\begin{table*}[htbp]
\begin{scriptsize}

    \centering
    \caption{Hardware and software environments}
    \label{tab:Hardware and software environments}
    \begin{tabular}{p{4cm} p{7cm} }

    \toprule \\
    \textbf{Hardware}\\ \\ 
    \toprule

   CPU &	Intel(R) Core(TM) i7-7660U CPU @ 2.50GHz 2.50 GHz \\
    
   Memory & 8.00 GB \\
   
   Hard Disk & 256 GB, SSD \\
   
   \bottomrule \\

    \textbf{Software}\\ \\
    \toprule
    
       OS & Linux 64-bit Ubuntu 20.04 LTS \\

       Docker & v 20.10.14 \\

       Docker-compose & v 1.27.2 \\

       Node & v 14.4.0 \\

       Golang & v 1.13.6 \\

       Hyperledger Fabric & v 2.3.1 \\

       Fabric- ca & v 1.4.9 \\

       CouchDB & v 3.1.0 \\

       Postman & v 9.4\\

       Visual Studio Code & v 1.63 \\

       Hyperledger Caliper & v 0.2.0 \\

    \bottomrule \\ 
\end{tabular}
    
\end{scriptsize}
\end{table*}

\begin{table*}[htbp]
\begin{scriptsize}

    \centering
    \caption{Docker images of nodes}
    \label{tab: Docker images of nodes}
    \begin{tabular}{p{4cm} p{2cm}}

    \toprule \\
    \textbf {Node type} & {Number}\\ \\ 
    \toprule

  	Peer node & 2\\
    
    Orderer node & 1\\
   
    Database node & 2\\

   CA node & 3\\

        Smart contract node & 2\\
   
   \bottomrule \\

\end{tabular}
    
\end{scriptsize}
\end{table*}

\section{Discussion and Limitations}

In this section, the Hyperledger Caliper tool is used to test the performance of some of the chaincode functions (i.e., AddPolicy, UpdatePolicy, QueryPolicy, DeletePolicy, and CheckAccess). Two different scenarios are applied to measure and compare the average latency and  throughput of the proposed DBC-ABAC model with the fabric-iot  \cite{ref7} and ABAC-HLFBC \cite{ref28}. The transaction latency can be expressed as the amount of time taken from submitting the transaction until the result becomes available through the whole network. The transaction throughput is generally defined as transactions per second (TPS), which is the rate required by the blockchain network to commit valid transactions in a specific period of time \cite{ref28}.

\begin{figure}[htp!]
   \centering
   \includegraphics[scale=0.3]{  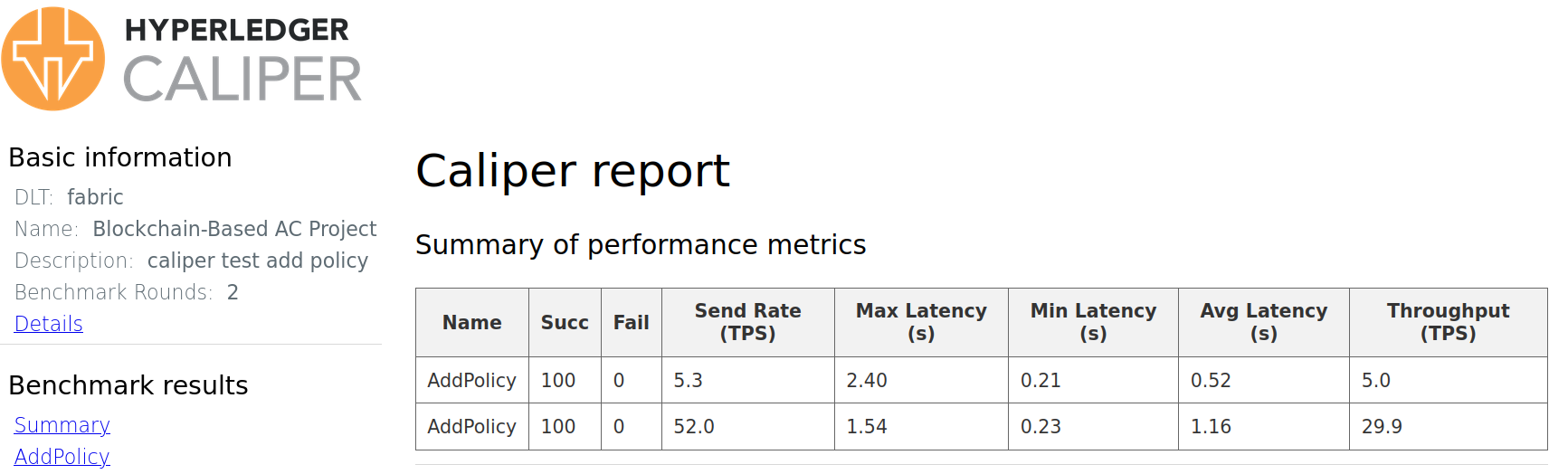}
    \caption{Caliper report of AddPolicy}
    \label{fig1:caliperReportAddPolicy}
\end{figure}

\begin{figure}[htp!]
   \centering
   \includegraphics[scale=0.3]{  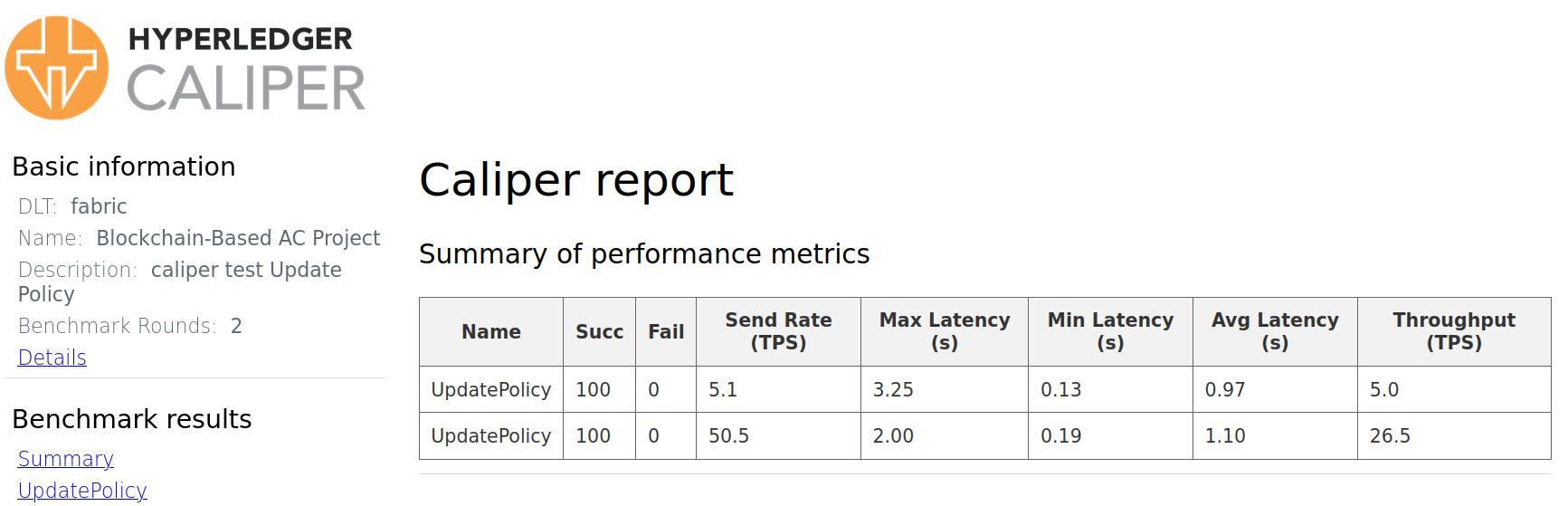}
    \caption{Caliper report of UpdatePolicy}
    \label{fig1:caliperReporUpdatePolicy}
\end{figure}

\begin{figure}[htp!]
   \centering
   \includegraphics[scale=0.3]{  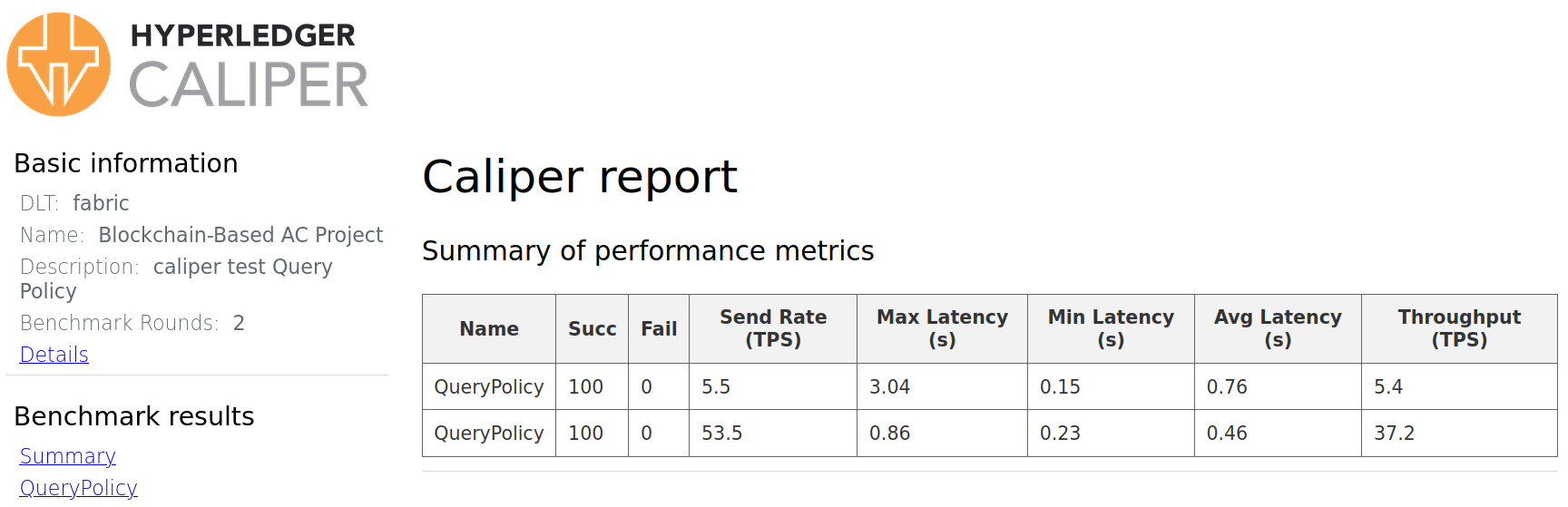}
    \caption{Caliper report of QueryPolicy}
    \label{fig1:caliperReporQueryPolicy}
\end{figure}

\begin{figure}[htp!]
   \centering
   \includegraphics[scale=0.3]{  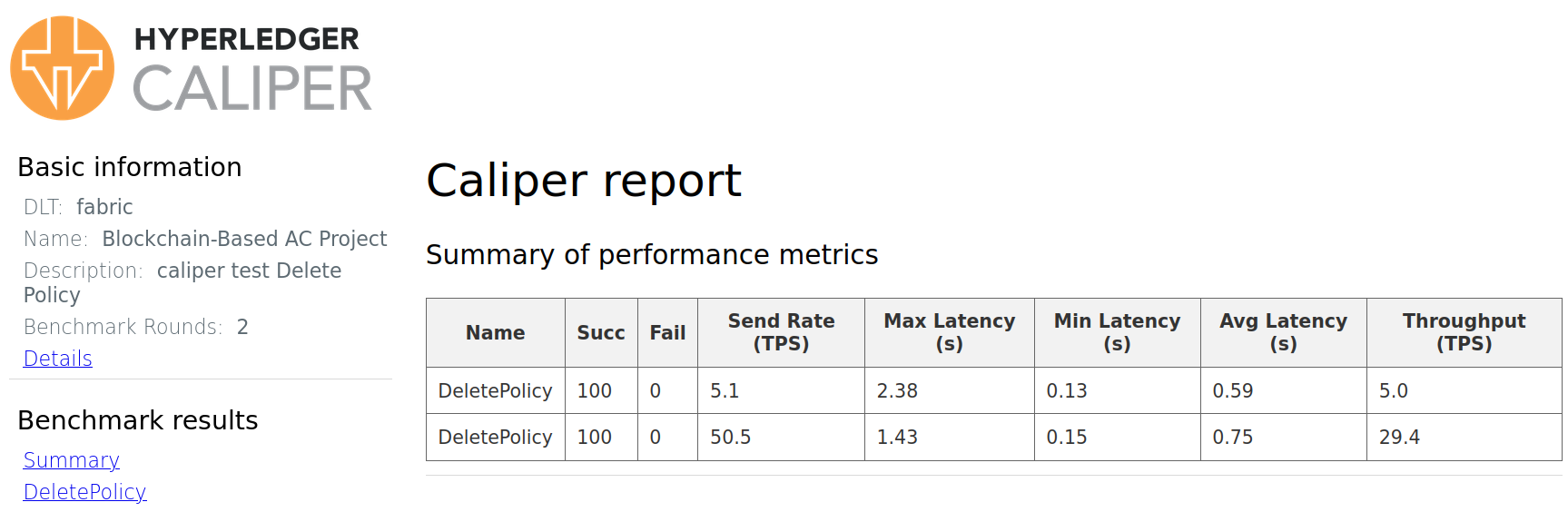}
    \caption{Caliper report of DeletePolicy}
    \label{fig1:caliperReporDeletePolicy}
\end{figure}

\begin{figure}[htp!]
   \centering
   \includegraphics[scale=0.3]{  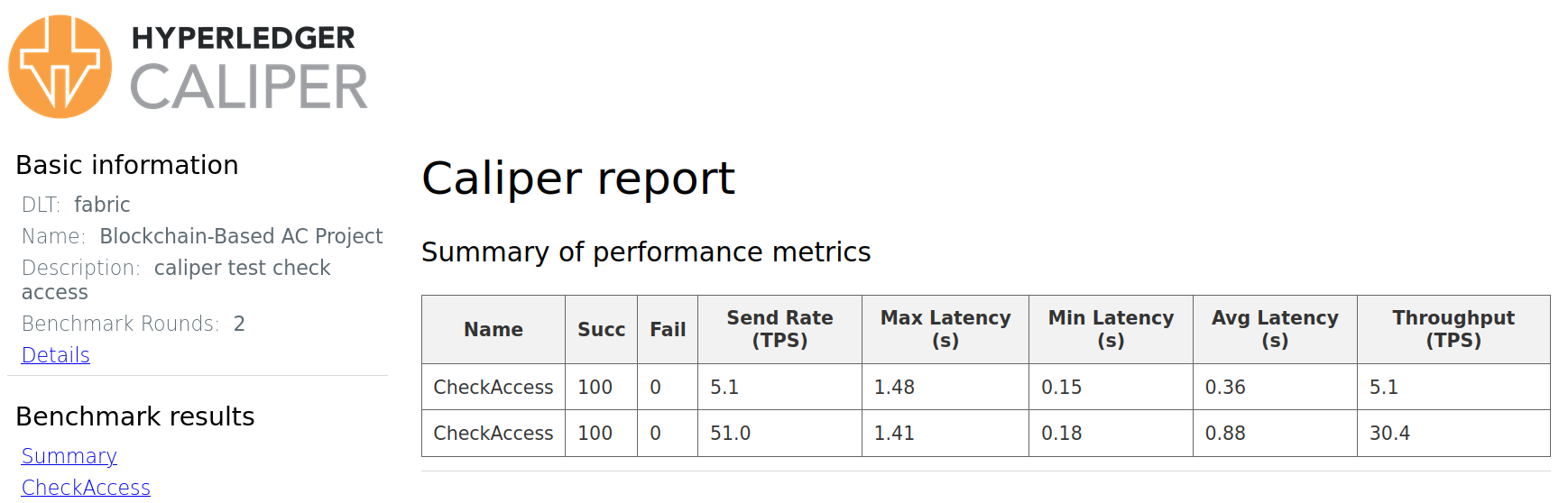}
    \caption{Caliper report of CheckAccess}
    \label{fig1:caliperReporCheckAccess}
\end{figure}

We set the send rate (TPS) to 5 in the first scenario and to 50 in the second one. The total number of transactions for both scenarios is 100. Figures ~\ref{fig1:caliperReportAddPolicy}, ~\ref{fig1:caliperReporUpdatePolicy}, ~\ref{fig1:caliperReporQueryPolicy}, ~\ref{fig1:caliperReporDeletePolicy}, and ~\ref{fig1:caliperReporCheckAccess} show the caliper reports of our AddPolicy, UpdatePolicy, QueryPolicy, DeletePolicy, and CheckAccess functions, respectively. \hfill \break

After measuring the average latency and throughput of the proposed model, we compare it with the fabric-iot  \cite{ref7} and ABAC-HLFBC \cite{ref28}.

\begin{figure}[htp!]
    \centering
    \includegraphics[scale=.42]{  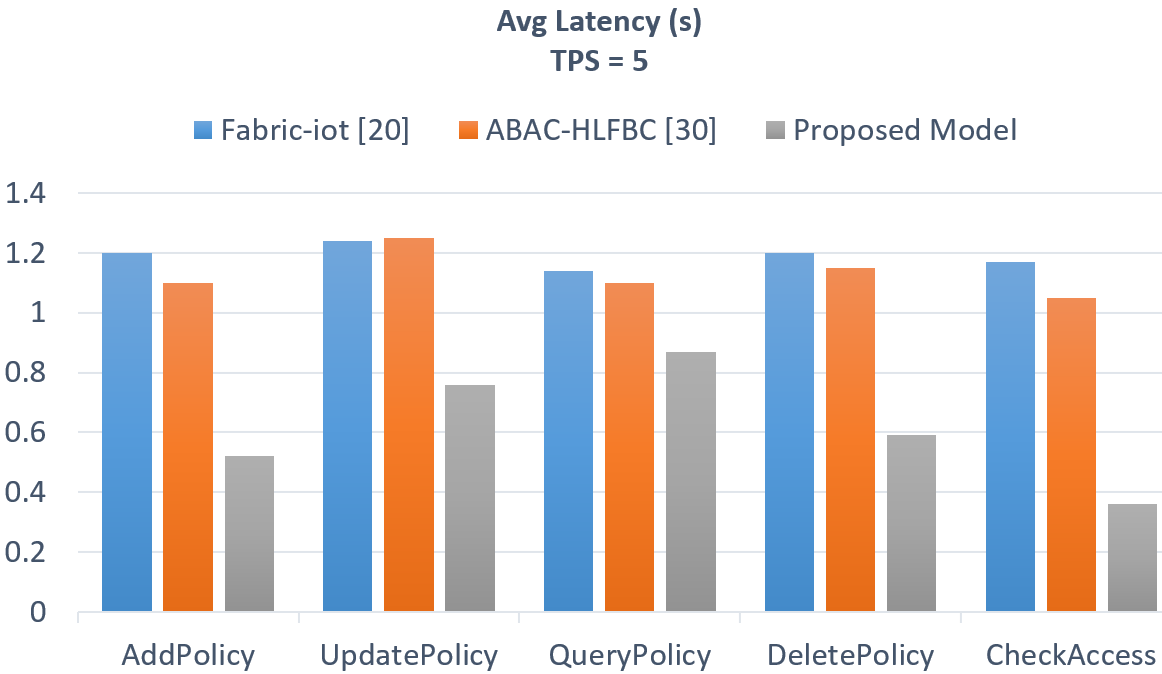}
    \caption{Comparison of the average latency when TPS = 5}
    \label{fig1:avgLatency_tps=5}
\end{figure}

\begin{figure}[htp!]
    \centering
    \includegraphics[scale=.42]{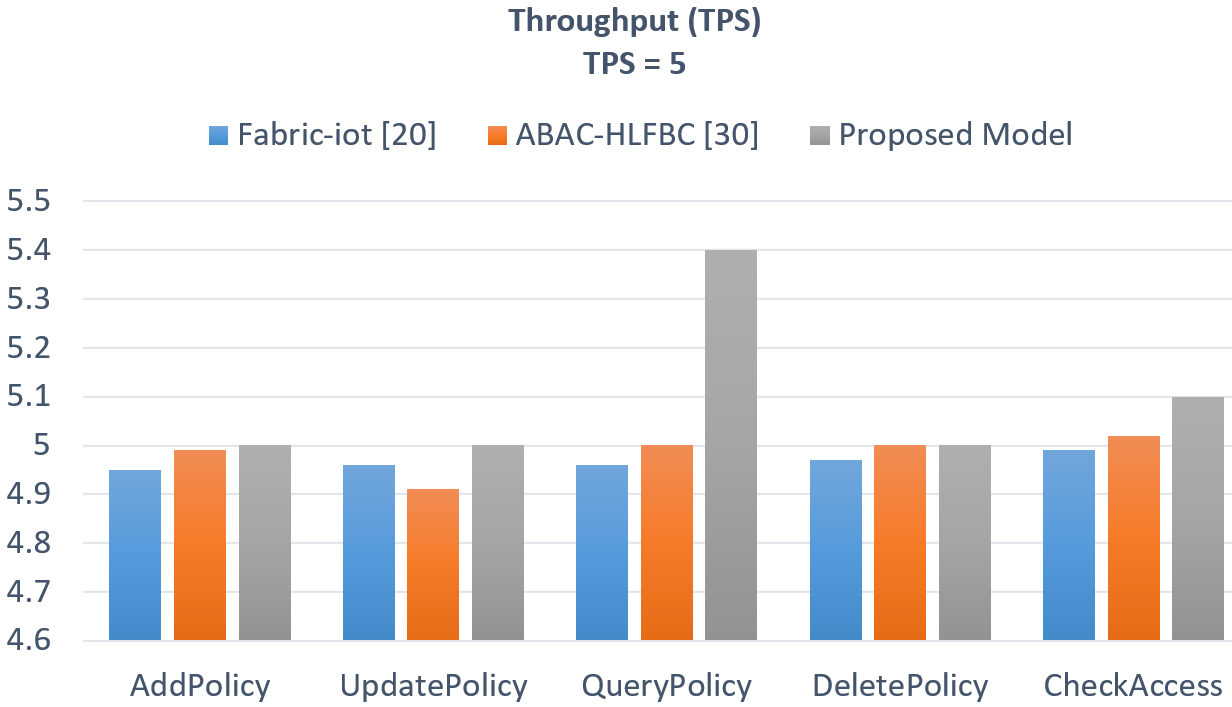}
    \caption{Comparison of the throughput when TPS = 5}
    \label{fig1:throughput_tps=5}
\end{figure}

Based on the first scenario, when TPS = 5, Figure ~\ref {fig1:avgLatency_tps=5} presents the comparison results of the average latency. The comparison shows that the average latency of the fabric-iot  and ABAC-HLFBC is more than one second for all smart contract functions. Whereas the proposed model minimizes the average latency to less than one second for all functions. Also, we find from Figure ~\ref{fig1:throughput_tps=5} that the throughput for the fabric-iot ranged from 4.95 to 4.99 and from 4.91 to 5.2 for ABAC-HLFBC. On the other hand, the throughput of our model varied between 5 and 5.4, which is almost better than the other models.

\begin{figure}[htp!]
    \centering
    \includegraphics[scale= .42]{  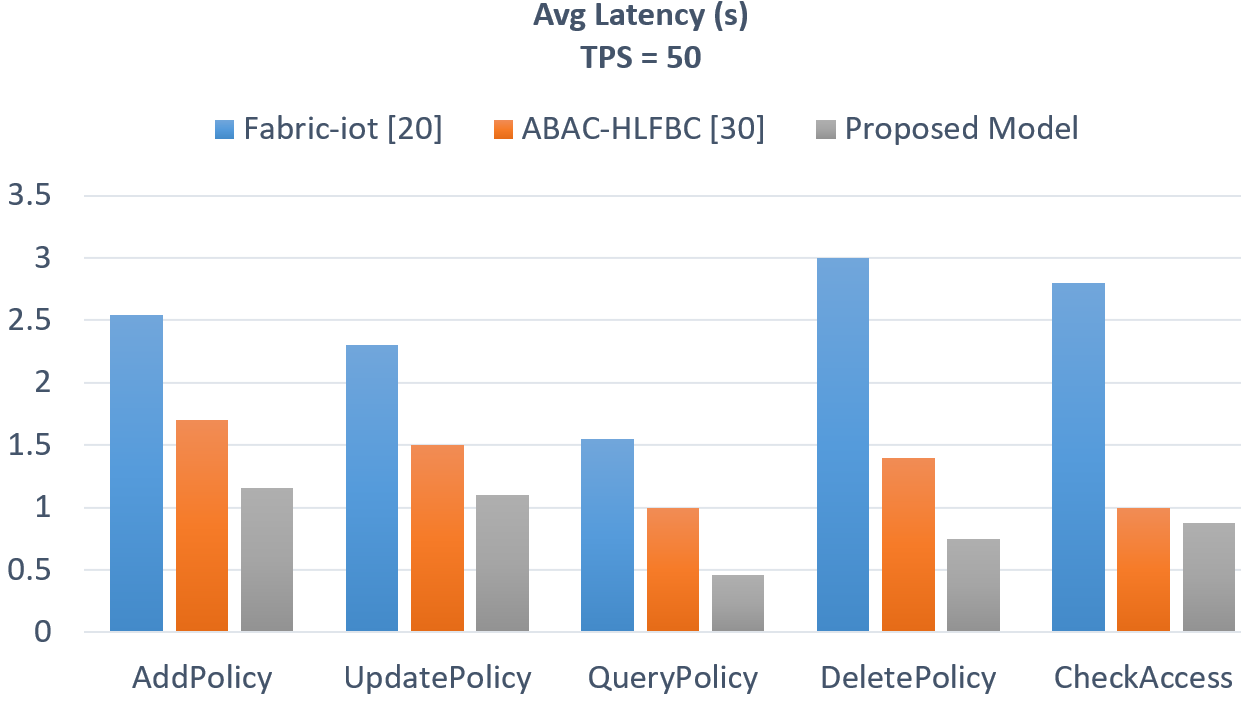}
    \caption{Comparison of the average latency when TPS = 50}
    \label{fig1:avgLatency_tps=50}
\end{figure}

\begin{figure}[htp!]
    \centering
    \includegraphics[scale=.35]{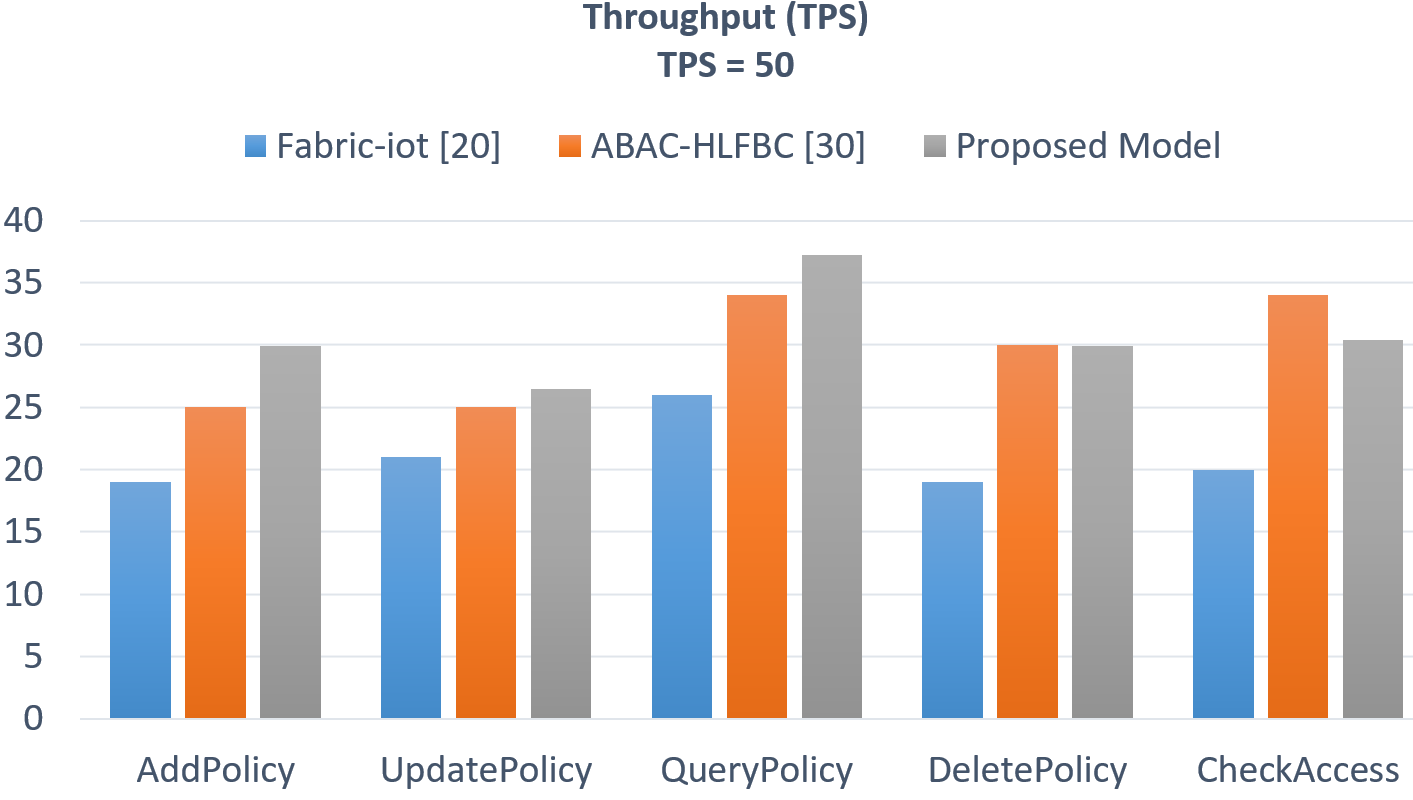}
    \caption{Comparison of the throughput when TPS = 50}
    \label{fig1:throughput_tps=50}
\end{figure}

For the second scenario, when TPS = 50, Figure ~\ref{fig1:avgLatency_tps=50} shows that the average latency of the proposed model is less than one second for QueryPolicy, DeletePolicy, and CheckAccess functions and around one second for AddPolicy and UpdatePolicy. Whereas the average latency for the fabric-iot exceeds two seconds for most functions and more than one second for the ABAC-HLFBC model. The representation of the caliper throughput report in Figure ~\ref{fig1:throughput_tps=50} shows that most of our chaincode functions (i.e., AddPolicy, UpdatePolicy, and QueryPolicy) achieve higher throughput. The throughput result for the DeletePolicy method shows no difference between the proposed model and ABAC-HLFBC, but it outperforms fabric-iot. For the CheckAccess chaincode operation, our work has less throughput result than the ABAC-HLFBC, but it outperforms the fabric-iot.

After measuring and comparing the latency and throughput of the proposed DBC-ABAC model, we compare it with the reviewed works in chapter 3 based on the predefined requirements. From Table ~\ref{tab:comparaison1}, we can see that our work achieves all the predefined requirements. The proposed DBC-ABAC model preserves lightweight in terms of communicational overhead due to the use of the ABAC model which is a lightweight access control model that does not require a significant amount of resources \cite{ref7}. Furthermore, the resource-constrained IoT devices are not deployed as blockchain nodes to avoid the computational overhead of these devices. To eliminate the communicational overhead, we apply a unicast transaction request between the blockchain nodes (i.e., edge device) rather than a broadcast request to retrieve the IoT data located at a different domain. The IoT gateway is also used as a bridge between the IoT devices and the blockchain to avoid the communication bottleneck caused by direct access by IoT devices to the blockchain. To avoid the storage limitation regarding the IoT devices and blockchain, the collected IoT data is stored in off-chain storage which is the distributed data storage system (DDSS). The proposed architecture achieves decentralization due to the use of distributed domains that are managed by distributed edge nodes to share IoT data through P2P communication. The collected IoT data is stored in DDSS rather than centralized storage (i.e., cloud), and the ABAC policies and components are distributed through blockchain ledger and smart contracts. For better scalability, we use the Hyperledger Fabric blockchain platform and Raft orderer service combined with the ABAC model to handle the growing number of IoT devices. Two smart contracts were designed to dynamically manage the access policies and enforce access decisions. We enhance the latency and throughput by combining edge computing with blockchain technology. Lastly, we support the delegation and revocation in the suggested model by simply changing the values of some ABAC attributes.

 \begin{table*}[h]
\begin{scriptsize}

\centering 
    \caption{Comparison of our proposed work and the existing works}
    \label{tab:comparaison1}
    \begin{tabular}{p{1.5cm}p{1.5cm} p{0.15cm} p{0.15cm} p{0.15cm} p{0.15cm} p{0.15cm} p{0.15cm} p{0.15cm} p{0.15cm} p{0.15cm} p{0.15cm} p{0.15cm} p{0.1cm}}
\multicolumn{2}{c}{Requirements} &	 

\multicolumn{1}{l}{\cite{ref10}}&	 
\multicolumn{1}{l}{\cite{ref20}} &
\multicolumn{1}{l}{\cite{ref5}}	&
\multicolumn{1}{l}{\cite{ref7}}	& 
\multicolumn{1}{l}{\cite{ref16}}&
\multicolumn{1}{l}{\cite{ref12}}&
\multicolumn{1}{l}{\cite{ref18}} &
\multicolumn{1}{l}{\cite{ref22}} &
\multicolumn{1}{l}{\cite{ref23}} &
\multicolumn{1}{l}{\cite{ref32}} &
\multicolumn{1}{l}{\cite{ref28}} &
\multicolumn{1}{l}{Ours}
\\ \toprule

\multirow{3}{*}{Lightweight}& Computation	 & $\times$	&\checkmark  & $\times$	 &	\checkmark  &	\checkmark  &$\times$	 &\checkmark &\checkmark & $\times$	  &\checkmark  &\checkmark &\checkmark \\ 

& Communication	&$\times$ 	&$\times$ &\checkmark	 &\checkmark	 &$\times$	 &\checkmark 	 &\checkmark &$\times$ &$\times$	 &\checkmark  &\checkmark &\checkmark\\ 

& Storage	&\checkmark &$\times$ &\checkmark	 &\checkmark	 &$\times$	 
&\checkmark 	 &\checkmark	&\checkmark &\checkmark  &\checkmark  &\checkmark &\checkmark\\  \hline

\multirow{2}{*}{Decentralization} &Architecture &$\times$ &\checkmark &$\times$  &\checkmark &$\times$ &$\times$ &\checkmark &$\times$ &\checkmark &\checkmark &\checkmark  &\checkmark\\ 

&Storage &$\times$ &\checkmark &$\times$ &$\times$ &\checkmark &$\times$ &$\times$  &\checkmark &$\times$  &\checkmark &$\times$ &\checkmark\\  \hline

\multicolumn{2}{l}{Scalability}	&\checkmark &\checkmark &$\times$	&\checkmark &$\times$ &\checkmark	 &\checkmark	 &$\times$ 	&\checkmark &\checkmark		&\checkmark &\checkmark \\

\multicolumn{2}{l}{Dynamicity}	&\checkmark	&\checkmark&	\checkmark&	\checkmark&	\checkmark&	\checkmark&	\checkmark &	\checkmark &	\checkmark &	\checkmark &	\checkmark &\checkmark\\

\multicolumn{2}{l}{Latency}	&$\times$ &\checkmark &$\times$ &\checkmark	&$\times$	&\checkmark	&$\times$ &$\times$ 	&\checkmark &\checkmark &\checkmark &\checkmark\\

\multicolumn{2}{l}{Throughput}&$\times$	&$\times$ &$\times$ &\checkmark &$\times$ &\checkmark &$\times$ &$\times$  &\checkmark &\checkmark &\checkmark &\checkmark\\

\multicolumn{2}{l}{Delegation} &$\times$	  &$\times$	  &$\times$	  &$\times$	  &\checkmark &\checkmark &$\times$ &$\times$	  &\checkmark &$\times$ &$\times$ &\checkmark\\

\multicolumn{2}{l}{Revocation}&	$\times$ 	&$\times$ 	&\checkmark&	\checkmark &$\times$	 &$\times$	 &	\checkmark &	\checkmark&	\checkmark&$\times$  &\checkmark	&\checkmark \\

\bottomrule
\end{tabular} 
    
\end{scriptsize}
\end{table*}

\section{Conclusion}\label{sec:conc}

The use of Internet of Things (IoT) technology has greatly enhanced many aspects of human life. To ensure the safety of sensitive data and to enhance security, access control is essential. Unfortunately, the centralized nature of traditional access control mechanisms is inadequate for the large-scale and distributed environment of the IoT.  As a result, there has been a growing interest in the use of blockchain technology to manage access control in the IoT, which is proving to be a promising solution

In this thesis, we conducted a comprehensive review of the most recent research related to the use of distributed blockchain-based access control for IoT systems. We proposed a set of important requirements that need to be met in order for the model to be successful. We then compared the reviewed works based on these requirements to  identify any gaps in the existing research. The analysis revealed that none of the researched models fully meet all the requirements.

To fill this research gap and achieve the predefined requirements, we propose a distributed blockchain-based access control model (DBC-ABAC)  that integrates blockchain technology with the attribute-based access control (ABAC) model to control access to IoT data. ABAC components are centralized in terms of attribute and policy storage and management. Accordingly, we bring these components (i.e., PDP, PIP, and PAP) to the blockchain, especially in the smart contract, to provide decentralized access control in distributed IoT environments. Also, in DBC-ABAC we combine the blockchain with the edge to bring the access decision operation near to the data producer  layer which reduces latency, enhances availability, and  eliminates the computing overhead for resource-constrained IoT devices. The ABAC  data access policy model is implemented using two smart contracts installed in each edge device (i.e., blockchain node) which are Policy Contract to manage access policies and Access Contract to provide access decisions to the data requester. 

To evaluate the performance of the proposed DBC-ABAC model, a proof-of-concept implementation was developed based on the Hyperledger Fabric blockchain platform.
The Hyperledger Caliper tool was integrated with Fabric   to measure and compare the transaction latency and throughput of various chaincode. The results indicated that the proposed model is capable of outperforming other works in terms of average latency and throughput.

The experiment conducted in this research was limited to a single local machine and did not consider the production environment, which could involve nodes distributed across different locations. Additionally, the consensus algorithm was not included in our scope. As a result, only one Raft ordering service was used for the implementation, which introduces potential centralization issues. Unfortunately, due to time constraints, the IoT part of the model and the distributed data storage system were not implemented

For future work, we are planning to Incorporate Multi-Level Security ABAC (MLS-ABAC), which is a lightweight scheme that provides data integrity and efficient access control and   extend the experiments by using more than one machine to verify the distributed performance of the proposed model.

 \bibliographystyle{agsm} 
 \bibliography{main}

@article{ref1,
  title={Blockchain in IoT: current trends, challenges, and future roadmap},
  author={Cui, Pinchen and Guin, Ujjwal and Skjellum, Anthony and Umphress, David},
  journal={Journal of Hardware and Systems Security},
  volume={3},
  number={4},
  pages={338--364},
  year={2019},
  publisher={Springer}
}

@article{ref2,
  title={A Review of Distributed Access Control for Blockchain Systems towards Securing the Internet of Things},
  author={Butun, Ismail and {\"O}sterberg, Patrik},
  journal={IEEE Access},
  year={2020},
  publisher={IEEE}
}

@article{ref3,
  title={Blockchain Platforms and Access Control Classification for IoT Systems},
  author={Abdi, Adam Ibrahim and Eassa, Fathy Elbouraey and Jambi, Kamal and Almarhabi, Khalid and AL-Ghamdi, Abdullah Saad AL},
  journal={Symmetry},
  volume={12},
  number={10},
  pages={1663},
  year={2020},
  publisher={Multidisciplinary Digital Publishing Institute}
}

@inproceedings{ref4,
  title={A survey on Blockchain based access control for Internet of Things},
  author={Riabi, Imen and Ayed, Hella Kaffel Ben and Saidane, Leila Azzouz},
  booktitle={2019 15th International Wireless Communications \& Mobile Computing Conference (IWCMC)},
  pages={502--507},
  year={2019},
  organization={IEEE}
}

@article{ref5,
  title={BorderChain: Blockchain-Based Access Control Framework for the Internet of Things Endpoint},
  author={Oktian, Yustus Eko and Lee, Sang-Gon},
  journal={IEEE Access},
  year={2020},
  publisher={IEEE}
}

@article{ref7,
  title={Fabric-IoT: A blockchain-based access control system in IoT},
  author={Liu, Han and Han, Dezhi and Li, Dun},
  journal={IEEE Access},
  volume={8},
  pages={18207--18218},
  year={2020},
  publisher={IEEE}
}

@article{ref9,
  title={LSB: A Lightweight Scalable Blockchain for IoT security and anonymity},
  author={Dorri, Ali and Kanhere, Salil S and Jurdak, Raja and Gauravaram, Praveen},
  journal={Journal of Parallel and Distributed Computing},
  volume={134},
  pages={180--197},
  year={2019},
  publisher={Elsevier}
}

@article{ref10,
  title={LBAC: A lightweight blockchain-based access control scheme for the internet of things},
  author={Qin, Xuanmei and Huang, Yongfeng and Yang, Zhen and Li, Xing},
  journal={Information Sciences},
  volume={554},
  pages={222--235},
  year={2021},
  publisher={Elsevier}
}

@article{ref12,
  title={BDSS-FA: A blockchain-based data security sharing platform with fine-grained access control},
  author={Xu, Hong and He, Qian and Li, Xuecong and Jiang, Bingcheng and Qin, Kuangyu},
  journal={IEEE Access},
  volume={8},
  pages={87552--87561},
  year={2020},
  publisher={IEEE}
}

@inproceedings{ref13,
  title={Hyperledger fabric: a distributed operating system for permissioned blockchains},
  author={Androulaki, Elli and Barger, Artem and Bortnikov, Vita and Cachin, Christian and Christidis, Konstantinos and De Caro, Angelo and Enyeart, David and Ferris, Christopher and Laventman, Gennady and Manevich, Yacov and others},
  booktitle={Proceedings of the thirteenth EuroSys conference},
  pages={1--15},
  year={2018}
}

@article{ref16,
  title={On the integration of blockchain to the internet of things for enabling access right delegation},
  author={Pal, Shantanu and Rabehaja, Tahiry and Hill, Ambrose and Hitchens, Michael and Varadharajan, Vijay},
  journal={IEEE Internet of Things Journal},
  volume={7},
  number={4},
  pages={2630--2639},
  year={2019},
  publisher={IEEE}
}

@article{ref17,
  title={Blockchain Consensuses Algorithms: A Survey},
  author={Ferdous, Md Sadek and Chowdhury, Mohammad Jabed Morshed and Hoque, Mohammad A and Colman, Alan},
  journal={arXiv preprint arXiv:2001.07091},
  year={2020}
}

@article{ref18,
  title={A novel attribute-based access control scheme using blockchain for IoT},
  author={Ding, Sheng and Cao, Jin and Li, Chen and Fan, Kai and Li, Hui},
  journal={IEEE Access},
  volume={7},
  pages={38431--38441},
  year={2019},
  publisher={IEEE}
}

@article{ref20,
  title={Scalable and secure access control policy for healthcare system using blockchain and enhanced Bell--LaPadula model},
  author={Kumar, Randhir and Tripathi, Rakesh},
  journal={Journal of Ambient Intelligence and Humanized Computing},
  volume={12},
  number={2},
  pages={2321--2338},
  year={2021},
  publisher={Springer}
}

@article{ref22,
  title={Capability-based iot access control using blockchain},
  author={Liu, Yue and Lu, Qinghua and Chen, Shiping and Qu, Qiang and O’Connor, Hugo and Choo, Kim-Kwang Raymond and Zhang, He},
  journal={Digital Communications and Networks},
  year={2020},
  publisher={Elsevier}
}

@article{ref23,
  title={xDBAuth: Blockchain based cross domain authentication and authorization framework for Internet of Things},
  author={Ali, Gauhar and Ahmad, Naveed and Cao, Yue and Khan, Shahzad and Cruickshank, Haitham and Qazi, Ejaz Ali and Ali, Azaz},
  journal={IEEE Access},
  volume={8},
  pages={58800--58816},
  year={2020},
  publisher={IEEE}
}

@article{ref28,
  title={An Attribute-Based Access Control Model for Internet of Things Using Hyperledger Fabric Blockchain},
  author={Shammar, Elham A and Zahary, Ammar T and Al-Shargabi, Asma A},
  journal={Wireless Communications and Mobile Computing},
  volume={2022},
  year={2022},
  publisher={Hindawi}
}

@inproceedings{ref29,
  title={An attribute based access control framework for healthcare system},
  author={Afshar, Majid and Samet, Saeed and Hu, Ting},
  booktitle={Journal of physics: conference series},
  volume={933},
  number={1},
  pages={012020},
  year={2017},
  organization={IOP Publishing}
}

@inproceedings{ref30,
  title={Access control for electronic health records with hybrid blockchain-edge architecture},
  author={Guo, Hao and Li, Wanxin and Nejad, Mark and Shen, Chien-Chung},
  booktitle={2019 IEEE International Conference on Blockchain (Blockchain)},
  pages={44--51},
  year={2019},
  organization={IEEE}
}

@inproceedings{ref31,
  title={An attribute based access control framework for healthcare system},
  author={Afshar, Majid and Samet, Saeed and Hu, Ting},
  booktitle={Journal of physics: conference series},
  volume={933},
  number={1},
  pages={012020},
  year={2017},
  organization={IOP Publishing}
}

@article{ref32,
  title={BEdgeHealth: A decentralized architecture for edge-based IoMT networks using blockchain},
  author={Nguyen, Dinh C and Pathirana, Pubudu N and Ding, Ming and Seneviratne, Aruna},
  journal={IEEE Internet of Things Journal},
  volume={8},
  number={14},
  pages={11743--11757},
  year={2021},
  publisher={IEEE}
}

@article{ref33,
  title={Decentralized Consensus Blockchain and IPFS-Based Data Aggregation for Efficient Data Storage Scheme},
  author={Subathra, G and Antonidoss, A and Singh, Bhupesh Kumar},
  journal={Security and Communication Networks},
  volume={2022},
  year={2022},
  publisher={Hindawi}
}

@article{ref34,
  title={Privacy preservation in blockchain based IoT systems: Integration issues, prospects, challenges, and future research directions},
  author={Hassan, Muneeb Ul and Rehmani, Mubashir Husain and Chen, Jinjun},
  journal={Future Generation Computer Systems},
  volume={97},
  pages={512--529},
  year={2019},
  publisher={Elsevier}
}

@inproceedings{ref36,
  title={PITI: Protecting Internet of Things via Intrusion Detection System on Raspberry Pi},
  author={Visoottiviseth, Vasaka and Chutaporn, Gannasut and Kungvanruttana, Sorakrit and Paisarnduangjan, Jirapas},
  booktitle={2020 International Conference on Information and Communication Technology Convergence (ICTC)},
  pages={75--80},
  year={2020},
  organization={IEEE}
}

@misc{ref40,
 
  title = {{A blockchain platform for the enterprise — hyperledger-fabricdocs master documentation,}},
  howpublished = "\url{https://hyperledger-fabric.readthedocs.io/en/release-2.3/}",
  year = {2023}, 
 
}

@misc{ref41,
  author={Pavan Adhav},
  title = {{Attribute-based access control (ABAC) in Hyperledger fabric}},
  howpublished = "\url{https://rb.gy/y6ujfb}",
  year = {Jan 26, 2020}, 
 
}

@misc{ref42,

  title = {{Hyperledger Caliper}},
  howpublished = "\url{https://hyperledger.github.io/caliper/v0.3.2/fabric-config/}",
  year = {2022}, 
 
}

@INPROCEEDINGS{ebtihel,
  author={Abdulrahman, Ebtihal and Alshehri, Suhair and Cherif, Asma},
  booktitle={2021 IEEE Asia-Pacific Conference on Computer Science and Data Engineering (CSDE)}, 
  title={Blockchain-Based Access Control for the Internet of Things: A Survey}, 
  year={2021},
  volume={},
  number={},
  pages={1-6},
  doi={10.1109/CSDE53843.2021.9718468}}





\end{document}